\documentclass[%
reprint,
amsmath,amssymb,
aps,]{revtex4-1}

\usepackage{graphicx}
\usepackage{epsfig}

\usepackage{dcolumn}
\usepackage{bm}

\newcommand*{\hatH}{\mathcal{H}}

\begin{document}
\title{
Multicanonical Sampling of the Space of States of $\hatH(2,n)$-Vector Models
} 

\author{Yu.~A. Shevchenko$^{1}$}
\email{shevchenko.ya@dvfu.ru}
\author{A.~G. Makarov$^{1}$}
\email{makarov.ag@dvfu.ru}
\author{P.D. Andriushchenko$^{1}$}
\email{pitandmind@gmail.com}
\author{Konstantin Nefedev$^{1,2}$}
\email{nefedev.kv@dvfu.ru}
\affiliation{
$^1$School of Natural Sciences, Far Eastern Federal University, Vladivostok, Sukhanova 8, 690091, Russian Federation \\
$^2$Institute of Applied Mathematics, Far Eastern Branch, Russian Academy of Science, Vladivostok, Radio 7, 690041, Russian Federation \\
$^3$Department of Physics, Tokyo Metropolitan University, Hachioji, 
Tokyo 192-0397, Japan \\
}

\date{\today}

\begin{abstract}
Problems of temperature behavior of specific heat are solved by the entropy simulation method for Ising models on a simple square lattice and a square spin ice (SSI) lattice with nearest neighbor interaction, models of hexagonal lattices with short-range (SR) dipole interaction, as well as with long-range (LR) dipole interaction and free boundary conditions, and models of spin quasilattices with finite interaction radius. It is established that systems of a finite number of Ising spins with LR dipole interaction can have unusual thermodynamic properties characterized by several specific-heat peaks in the absence of an external magnetic field. For a parallel multicanonical sampling method, optimal schemes are found empirically for partitioning the space of states into energy bands for Ising and SSI models, methods of concatenation and renormalization of histograms are discussed, and a flatness criterion of histograms is proposed. It is established that there is no phase transition in a model with nearest neighbor interaction on a hexagonal lattice, while the temperature behavior of specific heat exhibits singularity in the same model, in case of LR interaction. A spin quasilattice
is found that exhibits a nonzero value of residual entropy.
\end{abstract}

\maketitle

\section{Introduction}

Any equilibrium statistical properties of complex systems consisting of $N$ interacting particles can be calculated if the value of energy $E_i$ of each of $2^N$ states is known. The knowledge of the shape of the interaction energy distribution in the space of these states, $g(E)$, i.e., the distribution of the density of states (DOS) $g(E)/2^N$, or the so-called energy landscape \cite{Wales2003}, could allow one to precisely calculate thermodynamic averages. However, an exponential increase in the number of configurations and a complex growth of the number of allowed values of energy as a function of $N$ makes this problem extremely complicated. The equilibrium thermodynamics of a system of interacting magnetic moments can be analyzed by various approximate numerical methods. One of the most promising and actively developed methods that allows
the sampling of the space of states and calculate the $DOS$ is the method, proposed by Wang and Landau \cite{PhysRevLett.86.2050, Landau2004}, which is called the Wang–Landau (WL) method in western scientific literature. This method allows one to dynamically construct the $DOS$ by conditional random walks over the entire energy landscape by forming the statistical weights for degenerate or quasidegenerate (i.e., differing from each other by infinitely small interaction energy) states.

The wide spectrum of applied and fundamental problems, general function, immense field of applications, elegance, and obvious simplicity of the WL method have been the reason for the widest variety of its applications in statistical physics, biophysics, and other fields starting from spin systems \cite{PhysRevLett.86.2050, PhysRevE.64.056101,doi:10.1142/S0129183102003243,ref1,PhysRevLett.90.120201,PhysRevE.70.066128,Brown2005, PhysRevE.71.046705,PhysRevE.72.056710}, quantum systems \cite{PhysRevE.82.046703}, atomic clusters \cite{PhysRevE.89.013311,Calvo2003}, dipole \cite{PhysRevB.72.214203} and spin \cite{PhysRevLett.110.210603} glasses, liquid crystals \cite{PhysRevE.72.036702}, fluids \cite{Desgranges2009}, XY model \cite{Ngo2010}, the Blume–Capel model \cite{PhysRevE.92.022134}, the Potts model \cite{Caparica2015447}, biomolecules \cite{PhysRevE.90.042715}, protein folding \cite{Rathore2002}, polymer films \cite{TSAI2006}, as well as in many other fields of science, for example, in solving optimization problems \cite{ArgollodeMenezes2003428}, development of combinatorial number theory \cite{0305-4470-36-24-304}, and others.

The relatively fast convergence of the WL method makes it excellent for calculating the DOS; however, one should note that there are some questions on the accuracy of calculations, which have been pointed out by many authors (see, for example, \cite{PhysRevE.72.025701}). Certainly, the WL algorithm is one of the most interesting and unusual achievements of the last decade in Monte Carlo (MC) simulation methods \cite{PhysRevE.75.046701}. 

The method is based on the algorithm for calculating the DOS $g(E)$ — the relative number of possible states (configurations) for the energy level $E$. Thus, one can calculate any thermodynamic quantities that describe configurations arising during random walks over the space of states, including the free energy, in a wide range of temperatures.

The use of histograms for obtaining $g(E)$ significantly increased the level of simulation. This progress made it possible to obtain results for models of complex systems under test and allowed one to make significant progress in the numerical simulation and description of phase transitions. In contrast to the well-known and widely used MC algorithms, such as the Metropolis algorithm \cite{Metropolis1953}, the Swendsen–Wang cluster MC simulation \cite{PhysRevLett.58.86}, the parallel tempering or replica exchange MC algorithm  \cite{earl2005parallel}, the Wolff algorithm \cite{PhysRevLett.62.361}, and others, the WL algorithm allows one to obtain information on the canonical distribution $g(E) \exp\left[-E/k_B T\right]$ for given temperature $T$. The values of $g(E)$ for the same system may range from several units to several orders of magnitude. For example, in the Ising model, there are only two ground states on a square lattice of $100 \times 100$ spins; i.e.,  $g(E_{min}=-20000)=2$; however, in this case $g(E=0) \approx10^{3000}$.

The results of investigation of the efficiency and the convergence of the method \cite{Xu2015, Belardinelli2008, Maerzke2012}, as well as ways for its improvement \cite{PhysRevE.72.025701, Bornn2013749, Bauer2010}, are regularly published in the scientific literature. There also exist other constraints that have not yet been completely investigated, for example, a constraint on the speed of construction of flat histograms (see, for example, \cite{PhysRevLett.92.097201}). The renormalization scheme of the WL method was substantiated in \cite{PhysRevE.75.046701}. Among other problems formulated in \cite{PhysRevE.75.046701, Belardinelli2007}, which are very important for the practical application of the method, we can highlight the following. When a histogram can be considered as flat? What is the relationship between the value of the modification factor $f$ and the calculation error? Does there exist any universal method for controlling the convergence rate of the WL algorithm?

The energy space may have a very rough structure,
strong irregularities, steps, and forbidden gaps and discontinuities, whose neglect may have a significant
effect on the calculation error in thermodynamic
quantities. Such landscapes can be observed, for
example, in spin glasses and spin ice, which, owing to
their internal structure or lattice topology, can be
characterized by the absence or even unattainability of
equilibrium due, in particular, to the macroscopic
degeneracy of the ground state. Therefore, the application of the WL algorithm to the calculation of equilibrium thermodynamic properties of spin systems
that can pass to the state of spin glass or spin ice would
be very interesting. Note that, in the physics of phase
transitions, the problem of transition from the paramagnetic state to the spin glass state has not yet been
completely analyzed.

In the present study, we consider in detail serial and
parallel WL algorithms, try to answer the questions
posed above, and produce a solution to the problem of
calculation of the specific heat of the square Ising
model by the WL method to substantiate the efficiency
of the algorithm, a solution to the problem of the DOS
of SSI with a finite number of spins by a parallel WL
algorithm, a solution for fully connected models of
hexagonal spin lattices with long-range (LR) dipole–
dipole interaction, and a solution for spin quasilattices
with finite interaction radius.

\section{SERIAL METHOD OF CALCULATION}
\begin{figure}
\includegraphics[width=1\linewidth]{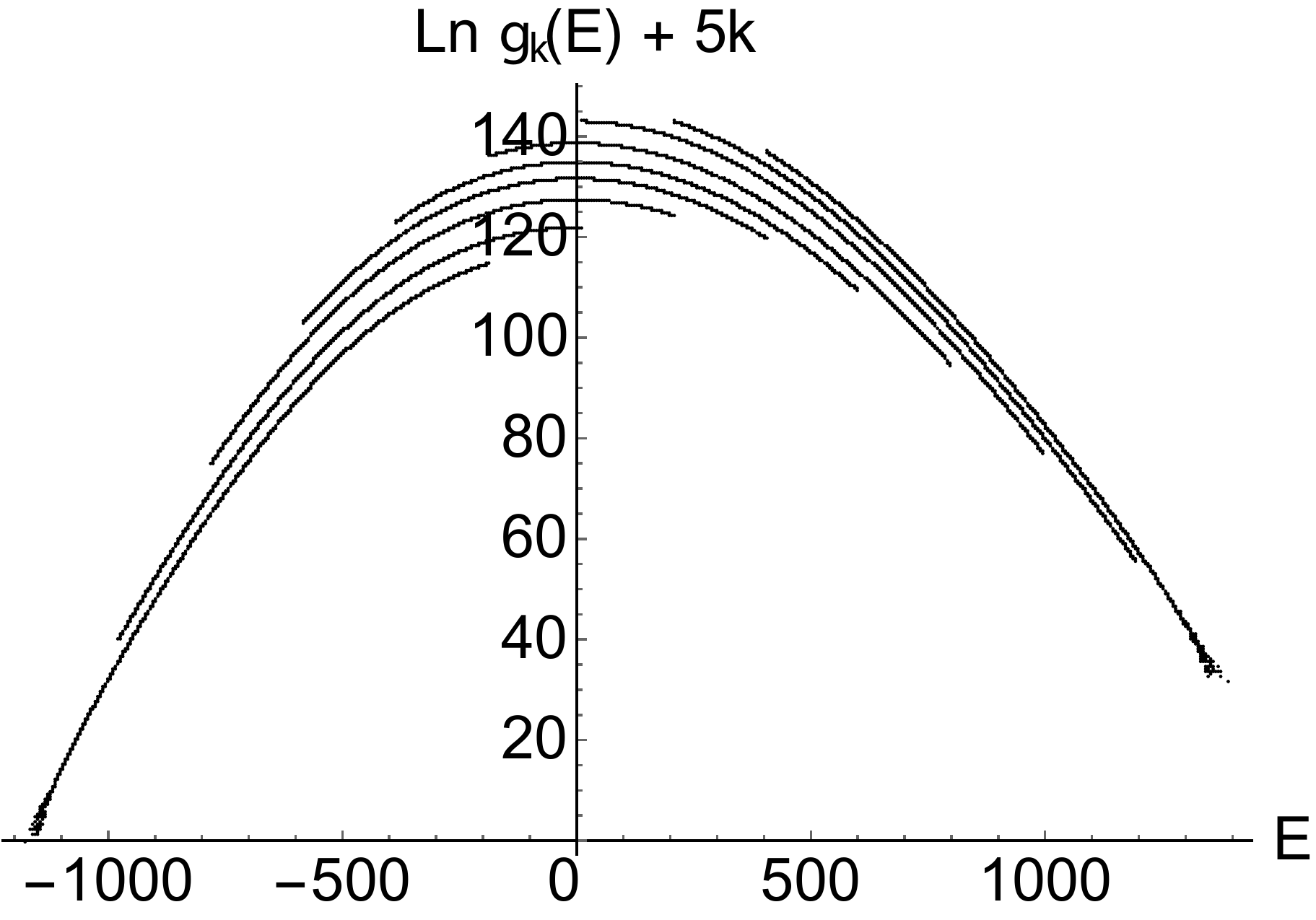}
\caption{Example of a normalized histogram $\ln[g(E)]$ obtained
for the SSI model with 180 spins by a parallel WL method
with LR interaction between dipoles. The energy space is
partitioned into nine equal intervals with 75\% overlapping.
The parts $g_k(E)$ are shifted along the ordinate for clarity.}
\label{g_e}
\end{figure}

Just as the Metropolis method  \cite{landau2000guide}, the WL
method belongs to the group of MC methods. These
methods are based on conditional random walks over
the space of states. However, the algorithms used in
these methods differ by the approach to sampling, i.e.,
by the method of constructing a sampled population,
a part of the universe of states, that is covered by a single experiment or by a series of serial or parallel experiments. The WL method uses an equiprobable sampling to construct the DOS. The probability distribution of energy levels, or the DOS, is represented as a
histogram $g(E)$, see Fig. \ref{g_e}.

In addition (see, for example, \cite{PhysRevE.84.065702,kalyan2016joint}), there exists a possibility to calculate a multidimensional distribution of the DOS of any thermodynamic quantity $X(E)$ measured in a numerical experiment and to obtain its mean value
\begin{equation}
\left<X\right>=\frac{\sum_E X(E) g(E)\exp\left[-\displaystyle\frac{E}{kT}\right]}{\sum_E g(E)\exp\left[-\displaystyle\frac{E}{kT}\right]}.
\label{XEGE}
\end{equation}
The histogram $g(E)$ is updated at every step irrespective of whether the configuration is accepted or rejected. This approach allows one to significantly reduce the operation time of the algorithm.

The operation principle and the details of implementation of a serial WL algorithm are described in the fundamental works \cite{PhysRevLett.86.2050, Landau2004}, see also \cite{silant2011, shchure2014}. In addition to $g(E)$, a histogram $H(E)$ is formed that serves as an indicator that all possible energy levels of the system are visited uniformly. The following initial values are defined:  $g(E)=1$, $H(E)=0$,  $\forall E$, the modification factor is $f=e^1\simeq2.718 281 8$; the effect of the accuracy of this factor on the result and the convergence rate will be discussed below.

The operation of the algorithm consists in step-by-step generation of a chain of states of the system. At every MC step, a candidate for a new configuration $C'_{i+1}$ is suggested that differs from the previous one by one flipped spin. With regard to the probability $P_{ac}$,
either the new configuration is accepted ($C_{i+1}= C'_{i+1}$), or the old one is returned $C_{i+1}= C_{i}$:

\begin{equation}
C_1 \xrightarrow{P_{ac}^{1,2}} C_2 \xrightarrow{P_{ac}^{2,3}} ... \xrightarrow{P_{ac}^{(n-1),n}}C_n.
\label{statesequence}
\end{equation}The updating probability
\begin{equation}
P_{ac}(E_{old}\rightarrow E_{new})=\min\left[ 1,\frac{g(E_{old})}{g(E_{new})}\right] 
\label{prob}
\end{equation}
depends on the probability distribution of energy states obtained at previous iterations of the algorithm, where $E_{old}$ and $E_{new}$ are energies corresponding to the old and new configurations.

The histograms are updated irrespective of whether
the configuration is accepted or rejected according to
the rule
\begin{equation}
\begin{cases}
g(E)\rightarrow g(E)*f\text{, where }f>1, \\
H(E)\rightarrow H(E)+1.
\end{cases}
\label{update_gh}
\end{equation}

In practice, in order to achieve the accuracy limits
guaranteed by the computer, it is desirable to use
$\ln g(E)$ instead of $g(E)$. Then the updating condition of
$g(E)$ is expressed as
\begin{equation}
\ln g(E) \rightarrow \ln g(E) +\ln (f).
\end{equation}
The sampling is performed until the histogram $H(E)$
becomes flat with certain accuracy. At this moment,
the WL step is terminated, and new values of $f=\sqrt{f}$ and $H(E)=0$, $\forall E$ are set. Thus, the algorithms starts a
new sampling cycle with higher updating accuracy of
$g(E)$. The accuracy of flatness is determined by the
maximum deviation of each element of the histogram $H(E)$ from its mean value. Usually, it is 80\%. According to the results presented in \cite{PhysRevLett.86.2050}, an increase in the threshold value deteriorates the convergence of the
algorithm. No increase in the accuracy of the result is
observed. The answer to the question of how the uniformity of the distribution of the auxiliary histogram
affects the accuracy of calculations requires additional
investigation. However, one can assert that an increase
in the uniformity of distribution leads to a decrease in
the efficiency of the algorithm.

A new distribution $g(E)$ is formed on the basis of
the previous one. The modification factor f simultaneously serves as a criterion of completion and an indicator of computing speed. The algorithm persists
while $f>f_{min}$, i.e., until a certain minimum value is
attained. Taking into account that every MC step
increases $g(E)$ by a factor f, by varying $f_{min}$, we actually
vary the accuracy of $g(E)$ due changing the number of
full WL cycles. This fact determines a balance between
the accuracy of $g(E)$ and the efficiency of the algorithm.

\section{PARALLEL SAMPLING}

The parallel WL method is a good symbiosis of two
MC algorithms: the serial WL method and the replica
exchange method, which is sometimes called a parallel
tempering method \cite{earl2005parallel}. In the classical WL algorithm,
a single serial process produces a random walk over the
whole space of states from $E_{min}$ to $E_{max}$. In the parallel
computation scheme \cite{PhysRevLett.110.210603,vogel2014scalable}, the energy space is
divided into $h$ overlapping intervals (energy windows
or bands). The value of overlap may vary depending on
the conditions of the problem \cite{PhysRevLett.110.210603,vogel2014scalable} or on the form
of the Gibbs distribution. Each energy interval is independently calculated by $m$ processes.

During operation, each process performs random
walks in its unique replica of the system and has individual independent distributions $g(E)$ and $H(E)$ and its own value of $f$.

During operation, processes $i$ and $j$ that occur in
adjacent windows and have their own respective histograms $g_i$ and $g_j$ exchange configurations $C_x$ and $C_y$ with
energies $E_x$ and $E_y$, respectively, with probability
\begin{equation}
P_{ex}=\min\left[1,\frac{g_i(E(C_x))}{g_i(E(C_y))}\frac{g_j(E(C_y))}{g_j(E(C_x))}\right].
\label{p_exchange}
\end{equation}

Note that a random $i$ process in one window
chooses for exchange a random $j$ process in one of two
adjacent windows. The exchange of configurations
between processes in windows that are not adjacent is
not admitted because, as a result of such an exchange,
energy would fall outside the limits of the energy window with greater probability.

Upon completing the exchange procedure, according to the rule (\ref{update_gh}), one should update the histograms
of each process involved in the exchange. To reduce
the calculation error, when a given value of the flatness
criterion of $H(E)$ is reached in all processes in one
energy window, $g(E)$ is averaged over all processes of
this window, and each process takes the average value
$\langle g(E)\rangle$   \cite{vogel2014scalable}. All values of the histogram $H(E)$ are set equal to zero.

\subsection{Energy Intervals}

To increase the efficiency of the method, one can
divide the space of states from $E_{min}$  to $E_{max}$  into equal
intervals with a given value of overlapping. Such
energy “bands” can be sampled independently. The
boundaries are defined as follows:
\begin{equation}
\Delta E = \frac{E_{max}-E_{min}}{h (1-l)+l},
\end{equation}
\begin{equation}
E^k_{min} = \Delta E(k-1)(1-l) + E_{min},
\end{equation}
\begin{equation}
E^k_{max} = \Delta E + E^k_{min},
\end{equation}
where $\Delta E$ -  is the width of the energy band, $E^k_{min}$ and $E^k_{max}$ are the lower and upper boundaries of the $k$th
interval, where the numbering of $k$ starts from $1$, and $l$
is the overlapping level of the intervals. For each
model, the size $\Delta E$ of the window is chosen empirically to obtain optimal computing time.

A nonoptimal value of $l$ may reduce the convergence rate of the algorithm but does not affect the
accuracy of calculations (except for the case of $l = 0$).
A too large value of $l$ leads to a significant increase in
the size of the window; as a consequence, a large number of states should be visited. For a small value of $l$,
the probability of failure during configuration
exchange noticeably increases. One of energies to be
exchanged may fall outside the energy window of the
process that accepts a configuration. The optimal
value of $l$ for the SSI model is $l = 0.8$.

In \cite{PhysRevLett.110.210603}, the authors additionally proposed to
increase $\Delta E$ dynamically and reduce $l$ as the energy
level increases. This approach is justified in the case of
systems with rough energy landscape. Near sharp
jumps in $g(E)$, one should reduce the values of $\Delta E$ and
$l$, while, in regions with smoother landscape, one
should increase the size of the window.

\subsection{Falling Outside the Energy Window}
During operation of the algorithm, it is critically
important to ensure that the values of energy obtained
satisfy the condition
\begin{equation}
E^k_{min} < E_i < E^k_{max},
\label{energy_range}
\end{equation}
Otherwise, problems associated with the return of the
process to its own energy window may arise. Before
starting the algorithm, one should balance the system,
i.e., make MC steps until each random process accepts
a replica whose energy falls in its own energy window.
For a known configuration of $E_{min}$, one should start
balancing from this configuration and make subsequent MC steps taking only those configurations that
increase energy.

When condition (\ref{energy_range}) is violated, one should cancel
the configurations ($P_{ac}=0$) during sampling. The
exchange of $C_x$ and $C_y$ configurations between processes $i$ and $j$ can occur only provided the following
conditions are satisfied:
\begin{equation}
\begin{cases}
E^i_{min}<E(C_y)<E^i_{max},\\
E^j_{min}<E(C_x)<E^j_{max}.
\end{cases}
\end{equation}
Otherwise the exchange is cancelled.

\subsection{Concatenation and Renormalization of Histograms}

A preliminary result of the parallel WL method is a
set of histograms $g_k(E)$ bounded by appropriate intervals $[E^k_{min},E^k_{max}]$, which should be concatenated into a resulting histogram $g(E)$. The histograms may significantly differ in height in different energy windows.

The process of concatenation consists in determining a gluing point followed by the normalization of each $k$th part to guarantee the continuity of $g(E)$. An
example of a normalized set $g_k(E)$ is shown in Fig. \ref{g_e}.
Two histograms are concatenated in a region where
their growth rates (angles of inclination) coincide, i.e.,
where the values of derivatives are maximally close to
each other.

\begin{figure}
\includegraphics[width=1\linewidth]{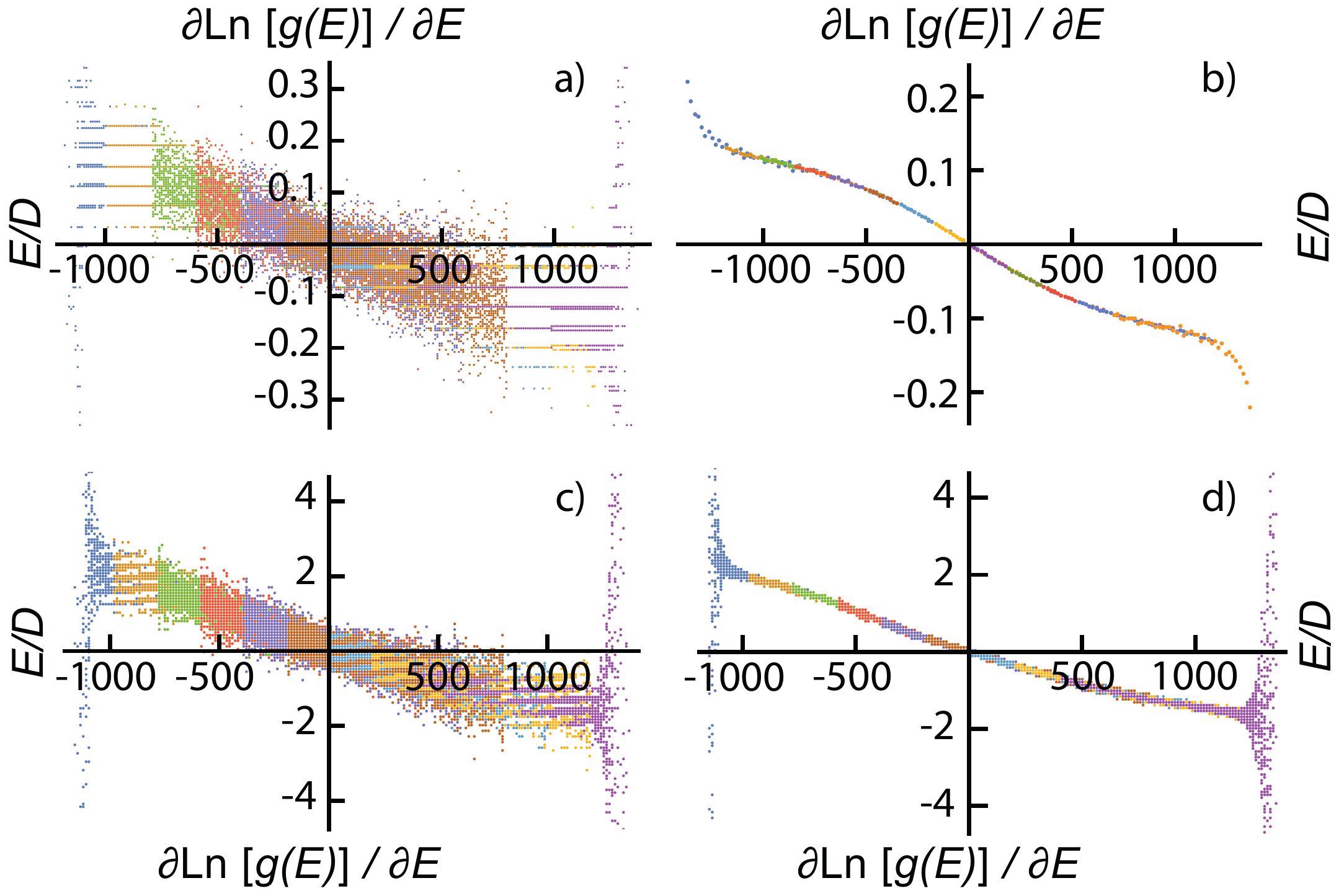}
\caption{
Examples of numerical differentiation of $g(E)$ by different methods. Two-point derivative of the histograms
of SSI with regard to (a) LR and (b) SR interactions. Differentiation of SSI histograms with regard to LR interaction obtained by
the five-point approximation method with variance of (c) 1 and (d) 10. Different energy intervals are marked by color.
}
\label{g_e_all}
\end{figure}

To study the models given below, we apply simple
two-point differentiation:
\begin{equation}
g'(E_i)=\frac{g(E_{i+1})-g(E_i)}{E_{i+1}-E_{i}}.
\end{equation}
This approach is effective in number series with small
variance, see Fig. \ref{g_e_all}b. However, it cannot be applied to
the analysis of systems with “dense” and “coarse”
energy landscape in view of the large dispersion of the
histogram values, see Fig. \ref{g_e_all}a. In such a case, one
should apply smoothing techniques before differentiation.

The authors of \cite{vogel2014scalable} propose using a five-point
approximation with given sequence length when the
distribution over states is smooth and has hardly any
forbidden gaps. The scope of approximation should be
large enough to obtain sufficiently smooth derivatives.
Examples of differentiation with scopes of 1 and 10 are
shown in Figs. \ref{g_e_all}c and \ref{g_e_all}d, respectively. In case of an
unsatisfactory result of matching the DOS regions,
one can apply additional line-smoothing techniques.

To calculate specific heat, one can use a relative
DOS; therefore, we calculated the specific heat of a
magnetic system on the basis of an unnormalized histogram $g(E)$ as a derivative of the internal energy:
\begin{equation}
C(T) = \frac { \partial \langle E(T)\rangle } { \partial T}.
\end{equation}

If one needs to calculate other thermodynamic
quantities, for example, entropy, one should obtain
true values of the degeneracy multiplicity of levels. In
the Ising ferromagnetic model, this can be done easily:
if the first element in the histogram $g(E)$ is $g(E_{min})=1$,
then all the other elements can be normalized by
$g(E_{min})$. In the Ising model, the ground state has multiplicity two; the same is observed in the SSI model
irrespective of the dipole–dipole interaction radius.

Since the parallel method of multicanonical sampling needs further development, all the resulted presented below were obtained by the serial WL method.

\section{RESULTS AND DISCUSSION}

\subsection{Critical Phenomena and Phase Transitions in $\hatH(d,n)$-Vector Models}

The divergence of thermodynamic quantities and
the stability or instability of thermodynamic behavior
for certain parameters (for example, the divergence of
specific heat and magnetic susceptibility for given values of the external magnetic field and temperature) are
usually called “critical phenomena” \cite{Ma1976}. They
accompany a second-order phase transition that
occurs at certain critical temperature $T_c$. It is assumed
that, in this transition, two phases are transformed
into each other without energy dissipation.

The continuous character of thermodynamic functions describing a second-order phase transition allows one to investigate the behavior of a system near $T_c$. It was even established that the behavior near the phase-transition temperature can be described by
power laws whose exponents depend on a very small number of parameters such as the dimension $d$ and the number $n$ of degrees of freedom of the system \cite{Fisher1974, Wilson1979,Fisher1998} (although the symmetry of the lattice and the interaction radius may change the character of critical behavior \cite{Fisher1974,Ma1976,Jose1977,Kosterlitz1978}). In fact, this means that the critical
behavior of most systems can be described by two
Hamiltonians involving microscopic pair interactions
\cite{Vaz2008}.

1. The Potts model of $Q$ states in which every $i$th
spin can be in one of Q possible discrete states (orientations) $\zeta_i$  $(\zeta_i = 1, 2, ..., Q)$. If two neighboring spins
have the same orientation, then they make a contribution $–J$ to the total energy of the configuration; otherwise, they make no contribution:
\begin{equation}
\hatH(d,n) = -J\displaystyle\sum_{\langle ij \rangle}\delta(\zeta_i, \zeta_j),
\end{equation}
where $\delta$  is the Kronecker delta $\langle ij \rangle$ denotes summation
over all interacting neighbors, and $J$ is the exchange constant.

2. The $n$-vector model, which is characterized by the fact that a spin can be in a continuum of states:
\begin{equation}
\hatH(d,n) = -J\displaystyle\sum_{\langle ij \rangle}{\vec S}_i\cdot {\vec S}_j,
\end{equation}
where ${\vec S}_i$ is an $n$-dimensional unit vector at site $i$ that interacts with vector ${\vec S}_j$ situated at cite $j$.

In this work, to study the thermodynamics of
phase transitions in complex spin systems by the WL
method considered above, we sampled spaces of
states of $\hatH(2,1)$- and $\hatH(2,2)$-vector models. The latter include the SSI model, the hexagonal spin ice
(HSI) model, and plane quasilattices of spin ice, the
so-called “spin snow.”

The Ising and SSI models were investigated by
short-range (SR) interaction; i.e., the interaction
between spins was taken into account only in the first
coordination sphere. The hexagonal model was calculated with both SR and LR interactions; the model
with LR interaction took into account interactions
between all spins.

Below we will show that, depending on the radius
and character of interaction, the same $\hatH(2,2)$-vector
model may exhibit a significant variation in the behavior of specific heat near $T_c$, which is accompanied by
the emergence of several maxima. We also show that,
in the case of HSI in the limit of an infinite number of
particles, the shape of a sample with free boundary
conditions does not affect the temperature behavior of
specific heat near $T_c$ irrespective of the interaction
radius.
 
\subsection{The Ising Model on a Square Lattice}

To check the accuracy of the WL algorithm, we
compared its results with the exact solution for the
simplest and well-studied Ising model on a simple
square lattice.

The Hamiltonian of the model is
\begin{equation}
\hatH(2,1) = -J\displaystyle\sum_{\langle ij \rangle}{S}_i\cdot {S}_j,
\end{equation}
where $S_i$ and $S_j$ take values $\pm 1$.

\begin{figure}[b]
\center{\includegraphics[width=1\linewidth]{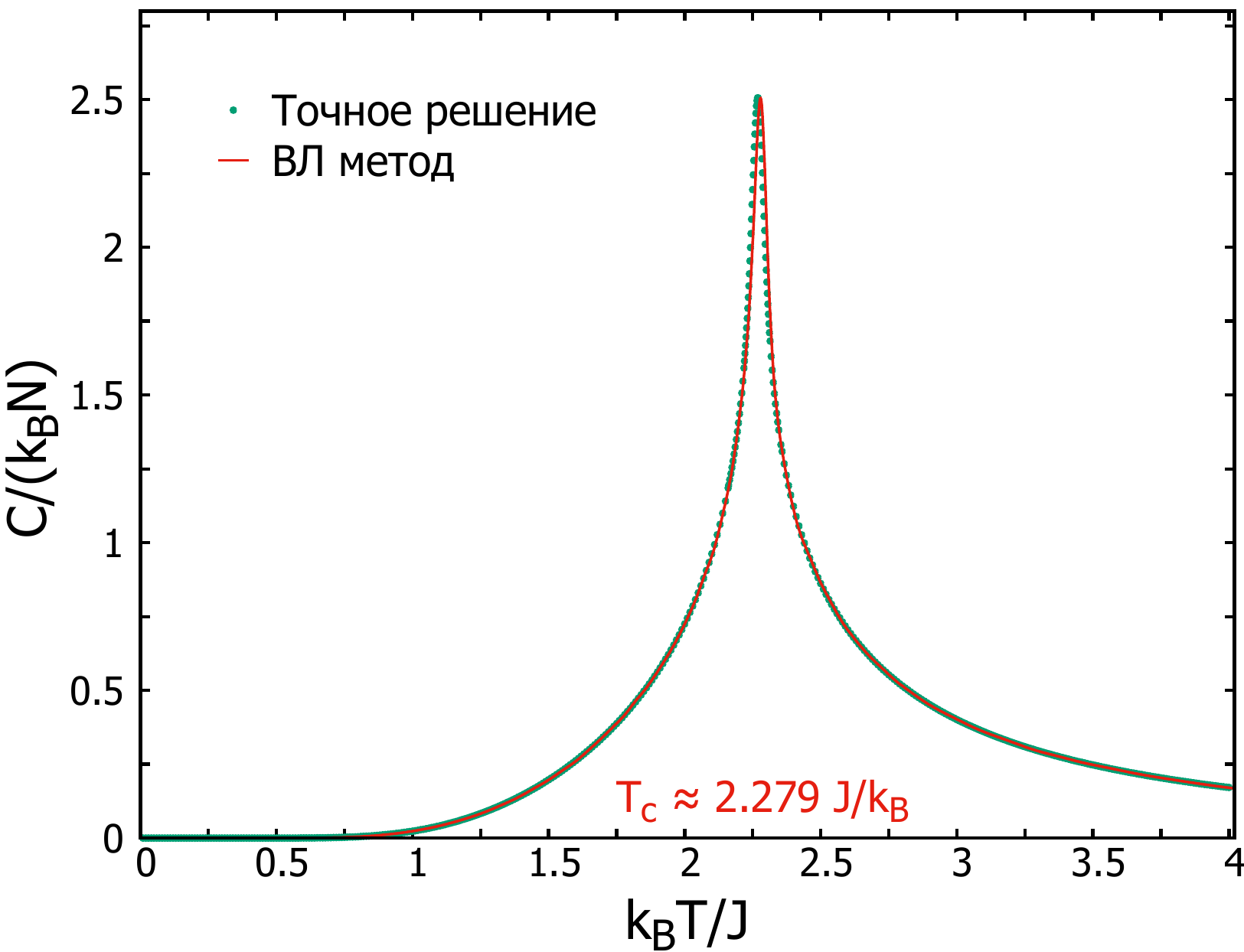}}
\caption{ 
Specific heat of a two-dimensional
Ising model of 104 spins obtained by two independent
methods: analytically \cite{Ferdinand1969} and by a serial WL method.}
\label{ising_wl}
\end{figure}

The size of the square lattice was $100 \times 100$ spins
with periodic boundary conditions. The flatness criterion of the histogram was $80\%$. We used 24 cycles of
WL sampling; i.e., at the last iteration, we had $f=e^{1/2^{23}}$. During operation, we performed altogether $~{3\times10^{11}}$ MC steps. The flatness of the histogram $H(E)$ was checked every $10^8$ steps.

Figure \ref{ising_wl} presents a comparison of the temperature
dependence of specific heat obtained by the WL
method and the exact analytic solution for $N=10^4$ \cite{Ferdinand1969}. The temperature of the peak was $T_c = 2.279$  $J/k_B$ for the WL method and  $T_c=2.269$  $J/k_B$  for the exact solution. The values of specific heat at the peak were
$2.50497$ and $2.50688$in dimensionless units, respectively. The relative deviation of specific heat from the
exact solution was less than $0.08\%$. Figure \ref{ising_wl} presents
the data of one run of the algorithm. The deviation
between independent runs of the algorithm was less
than the point size in the figure.

\subsection{Artificial Arrays of Superspin Ice}

The nanoarchitecture of artificial superspin (macrospin) ice represents an array of single-domain ferromagnetic nanoparticles of given geometry arranged on
a substrate or in the bulk of a nonmagnetic material.
The nanoparticles are usually made of thin-film magnetic materials, for example, permalloy or cobalt. The
volume of a nanoisland is on the order of $\sim10^5$ $nm^3$, and
the magnetization is on the order of $\sim10^{7} \mu_B$. The shape
anisotropy due to the geometry of the nanoisland leads
to the parallel ordering of atomic magnetic moments
along the long axis, while the behavior of the magnetic
moment of the nanoisland is Ising-like (superspin or
macrospin), as demonstrated in a number of experimental works \cite{wang2006artificial,PhysRevB.77.094418,Lederman1993, PhysRevB.80.140409}.

In \cite{Nefedev2010, Ivanov2011, Nefedev2011}, the authors presented the results of
numerical simulation of magnetic force microscopy
(MFM) experiments on the basis of which one can
conclude that the MFM images of nanoparticle arrays
(see, for example, \cite{RevModPhys.85.1473} do not contain non-singledomain states of nanoislands and that the deviation
from the uniform magnetization due to the presence of
edge effects can be neglected even for square nanoparticles. For rectangular thin-film islands with significant shape anisotropy, the edge effects are still weaker.
This give grounds for the application of the Ising spin
model with dipole interaction to describe the equilibrium thermodynamics of dipole ice.

The model of artificial dipole ice was presented in
our recent paper \cite{Shevchenko2017}. The Hamiltonian of dipole
interaction is
\begin{equation}
\hatH(2,2)=\sum_{\langle ij \rangle} E^{ij}_{dip},
\label{dipint1}
\end{equation}
and the energy of dipole–dipole interaction for an $ij$
pair of spins is
\begin{equation}
E^{ij}_{dip}=D \left(
\frac{
	({\vec m}_i{\vec m}_j)
}{
	\vert{\vec r}_{ij}\vert^3
}
-3
\frac{
	({\vec m}_i{\vec r}_{ij})
	({\vec m}_j{\vec r}_{ij})
}{
	\vert{\vec r}_{ij}\vert^5
}\right),
\label{dipint2}
\end{equation}
where  
$D=\mu^2/a^3$ is a dimensional coefficient,
$\mu$ is the total magnetic moment of the island, and $a$ is the lattice parameter. Each nanoisland is considered as an Ising-like magnetic dipole. 

All the solutions of the problems on the temperature behavior of specific heat presented in this paper
have been obtained within equilibrium thermodynamics, i.e., under the assumption that the energy barriers
controlled by the shape anisotropy or by other kinds of
anisotropies are overcome.

\subsubsection{Square spin ice}

\begin{figure}
\center{\includegraphics[width=0.7\linewidth]{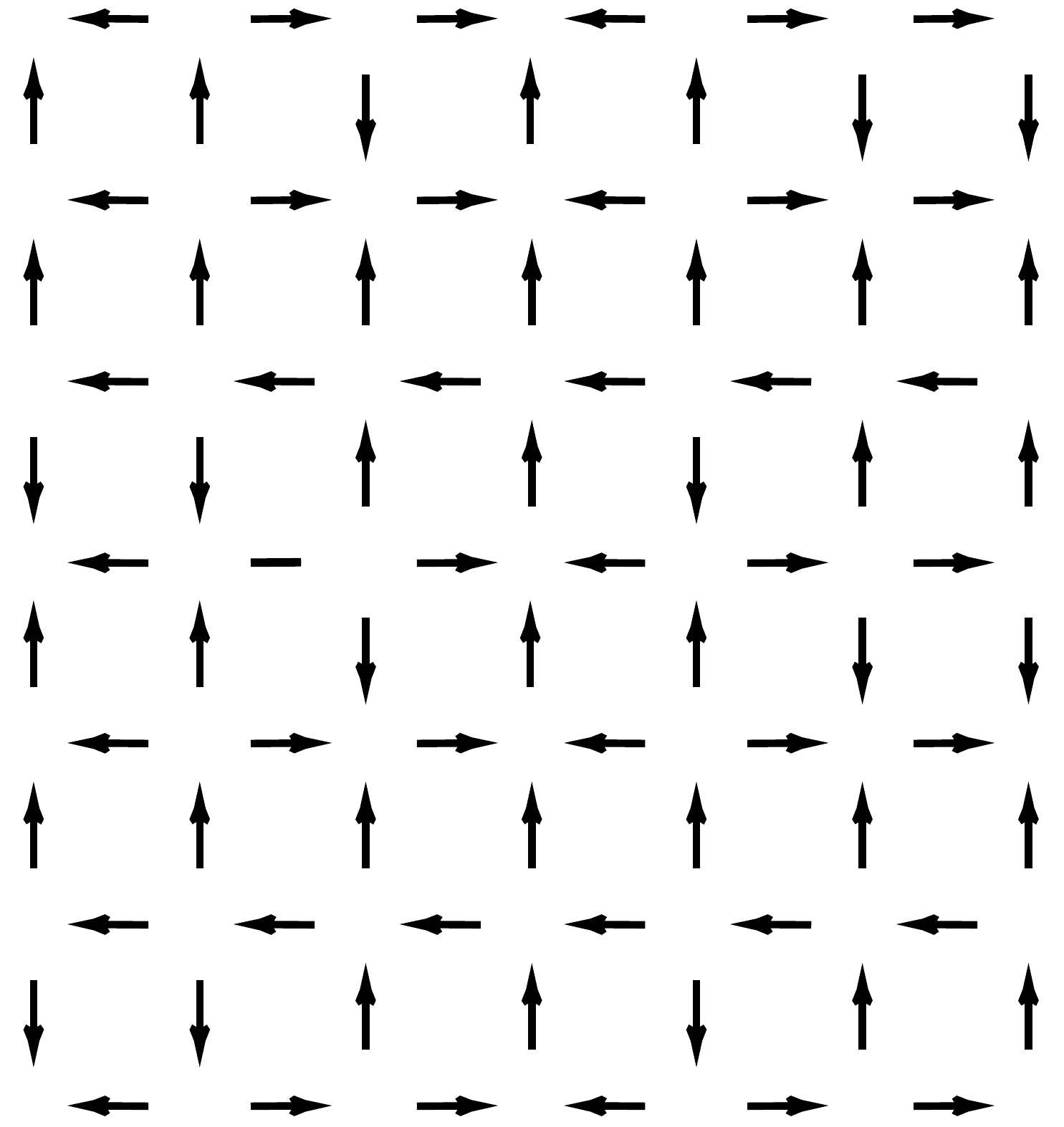}}
\caption{Schematic view of the square spin ice model.}
\label{SSI}
\end{figure}

Using a parallel WL method, we obtained the DOS for the SSI model (Fig. \ref{SSI}) with
a size of $70 \times 70$ unit cells (9940 dipoles) and SR interaction (four neighbors) with free boundary conditions.
The energy space was partitioned into 80 equal intervals with 80\% overlapping. The algorithm was run on
400 computer cores, with five replicas per interval.

The results obtained by the WL method are in good
quantitative and qualitative agreement with the results
obtained by the Metropolis algorithm to within experimental error (Fig. 5). The peak temperature is $T_c \approx 9.52 D/k_B$. However, this value is different from the
result $T_c \approx 7.2 D/k_B$ obtained in \cite{silva2012thermodynamics}. The difference in
temperature is attributed to the different number of interacting nearest neighbors used in the model. The
dipole–dipole interaction is of antiferromagnetic
character; therefore, it leads to a decrease in the temperature of phase transition with increasing number of
interacting neighbors in SSI.

\begin{figure}
\includegraphics[width=1\linewidth]{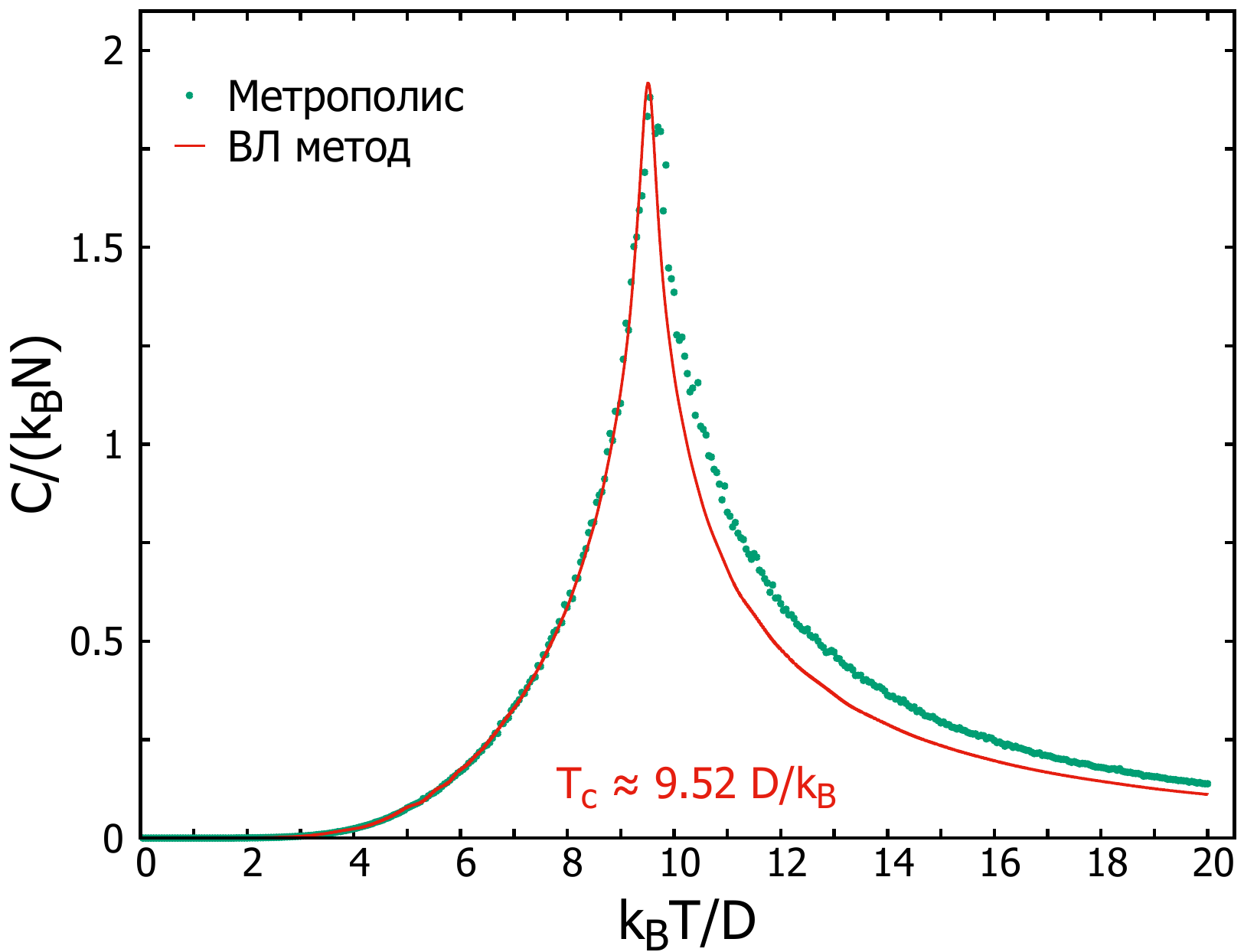}
\caption{Specific heat of the square spin ice
model (9940 dipoles) with nearest neighbor interaction
obtained by Metropolis methods (dots) and by a serial WL
method (solid line).}
\label{ssi_wl}
\end{figure}

\subsubsection{Hexagonal spin ice}

We investigated two
series of samples consisting of different numbers of
particles formed into arrays of hexagonal spin ice of
square and hexagonal (“from the center”) shape with
different interaction radii. In the SR case, the interaction only with four (or less, on the boundaries) nearest
neighbors was taken into account. In the LR case,
each spin interacted with each other. The thermodynamics of the samples was calculated in 24 steps of a
serial WL algorithm.

\begin{figure}
\center{\includegraphics[width=0.8\linewidth]{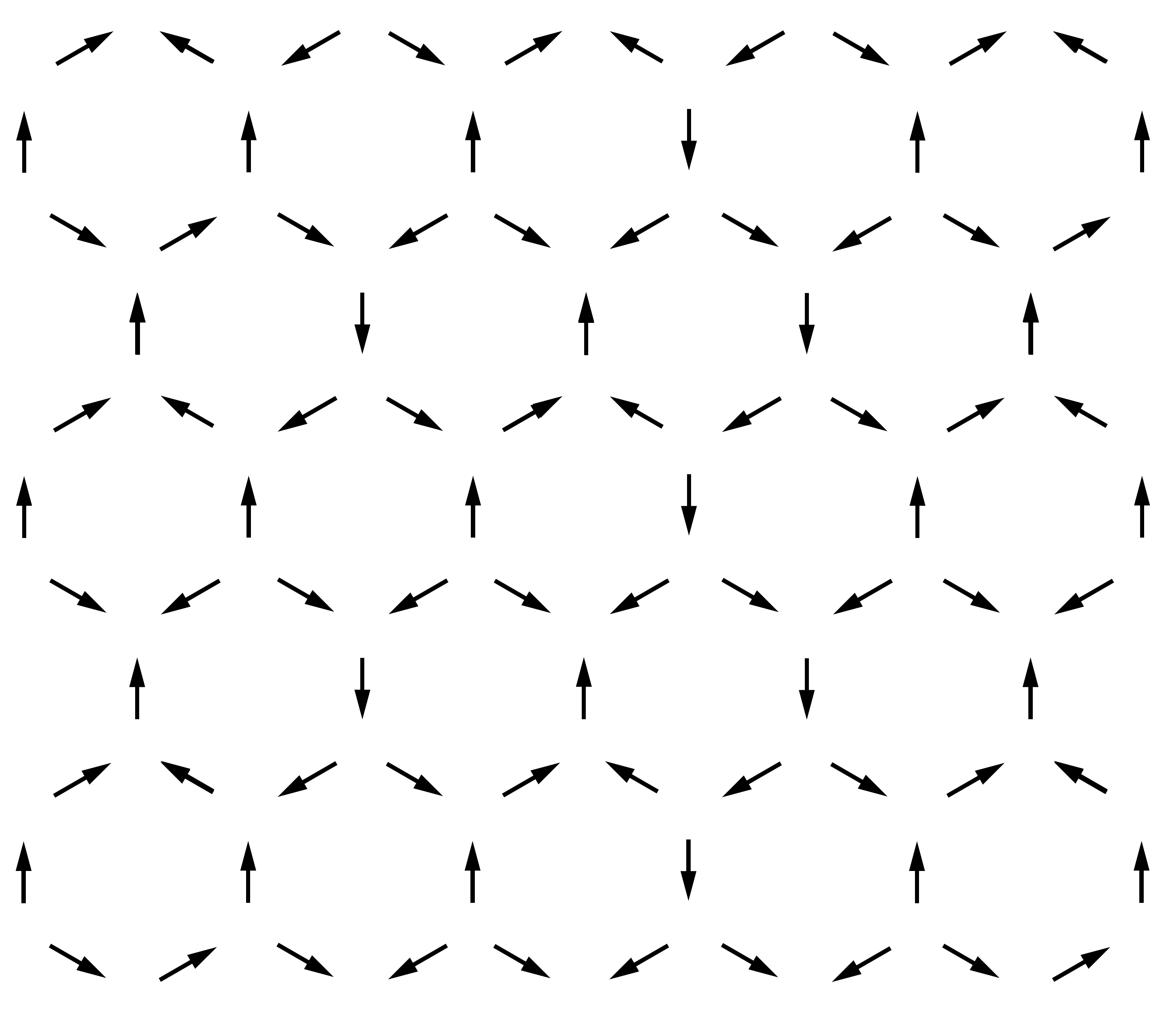}}
\caption{Schematic view of a square sample of hexagonal
spin ice.}
\label{Sq_sample}
\end{figure}

\begin{figure}
\includegraphics[width=1\linewidth]{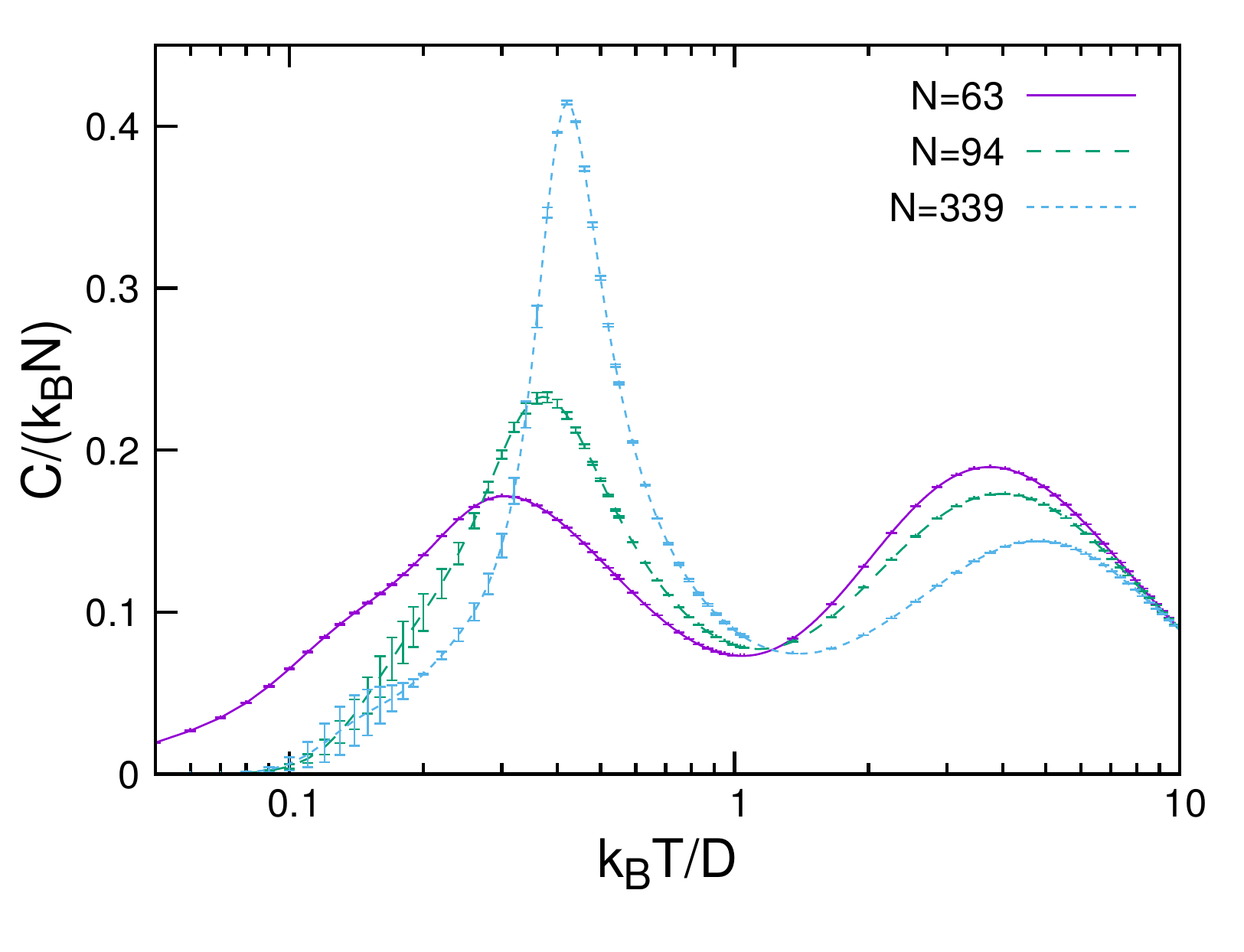}
\caption{Specific heat as a function of temperature for hexagonal spin ice (square sample) for LR
interaction. }
\label{Sq_LR}
\end{figure}

\begin{figure}
\includegraphics[width=1\linewidth]{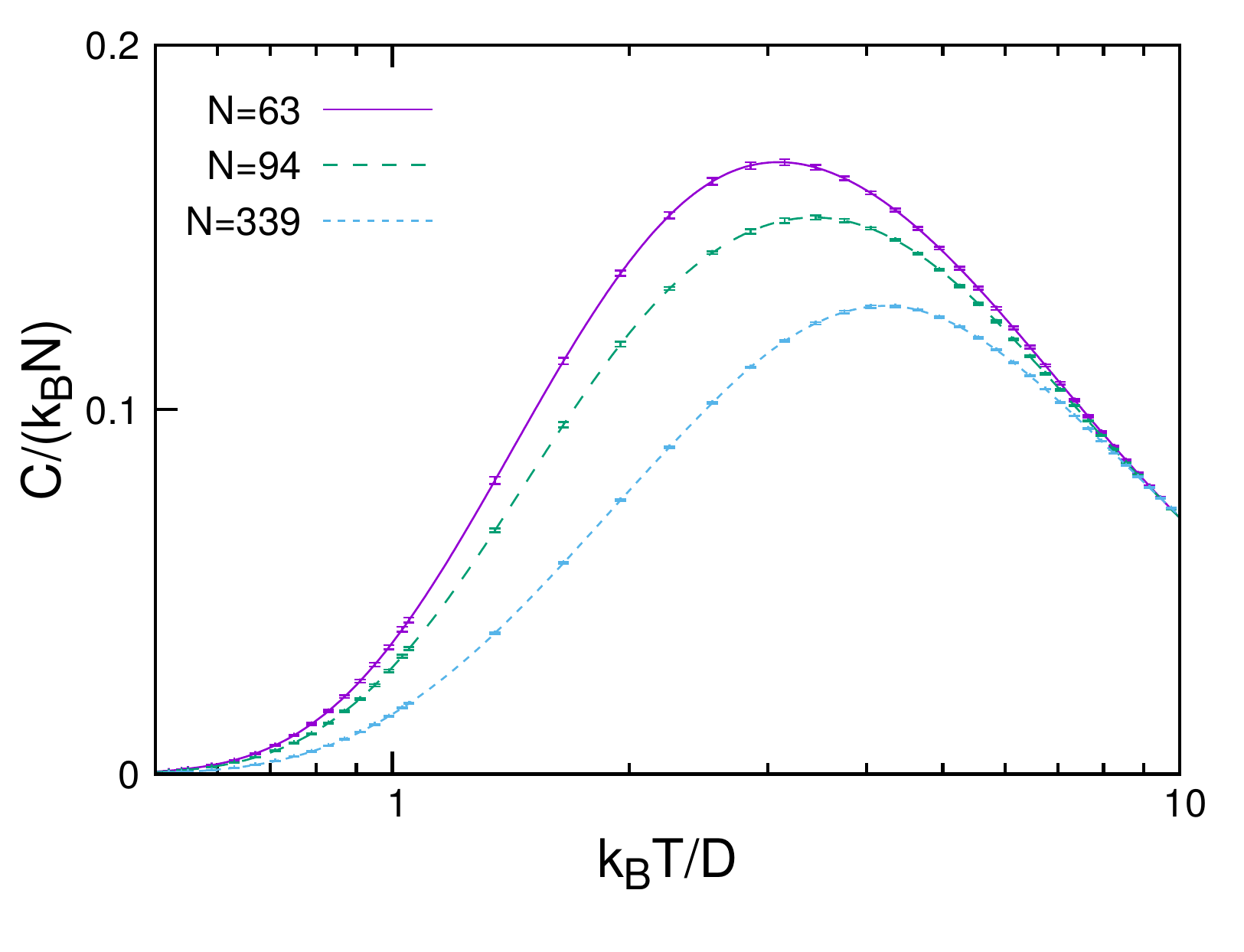}
\caption{Specific heat as a function of temperature for hexagonal spin ice (square sample) for SR
interaction.}
\label{Sq_SR}
\end{figure}

The model of spin ice in the form of a square sample of hexagonal lattice with free boundary conditions,
dipole interaction (\ref{dipint1}), and $5 \times 5$ cells, i.e., $N = 94$
magnetic moments, is schematically shown in Fig. \ref{Sq_sample}.
The temperature behavior of the specific heat of $4 \times 4$
($N = 63$), $5 \times 5$ ($N = 94$), and $10 \times 10$ ($N = 339$) models is demonstrated in Figs. \ref{Sq_LR} and \ref{Sq_SR}.

In case of LR interaction (Fig. \ref{Sq_LR}), the temperature
behavior of specific heat exhibits an anomalous character, and two temperature peaks are observed; i.e.,
the critical behavior of the system of dipoles is substantially changed near $T_c$. The first peak increases
with increasing number of particles, while the second
peak decreases. Figure  \ref{Sq_SR} shows that the specific heat
peak in models with SR interaction decreases with
increasing number of particles.

\begin{figure}
\center{\includegraphics[width=0.8\linewidth]{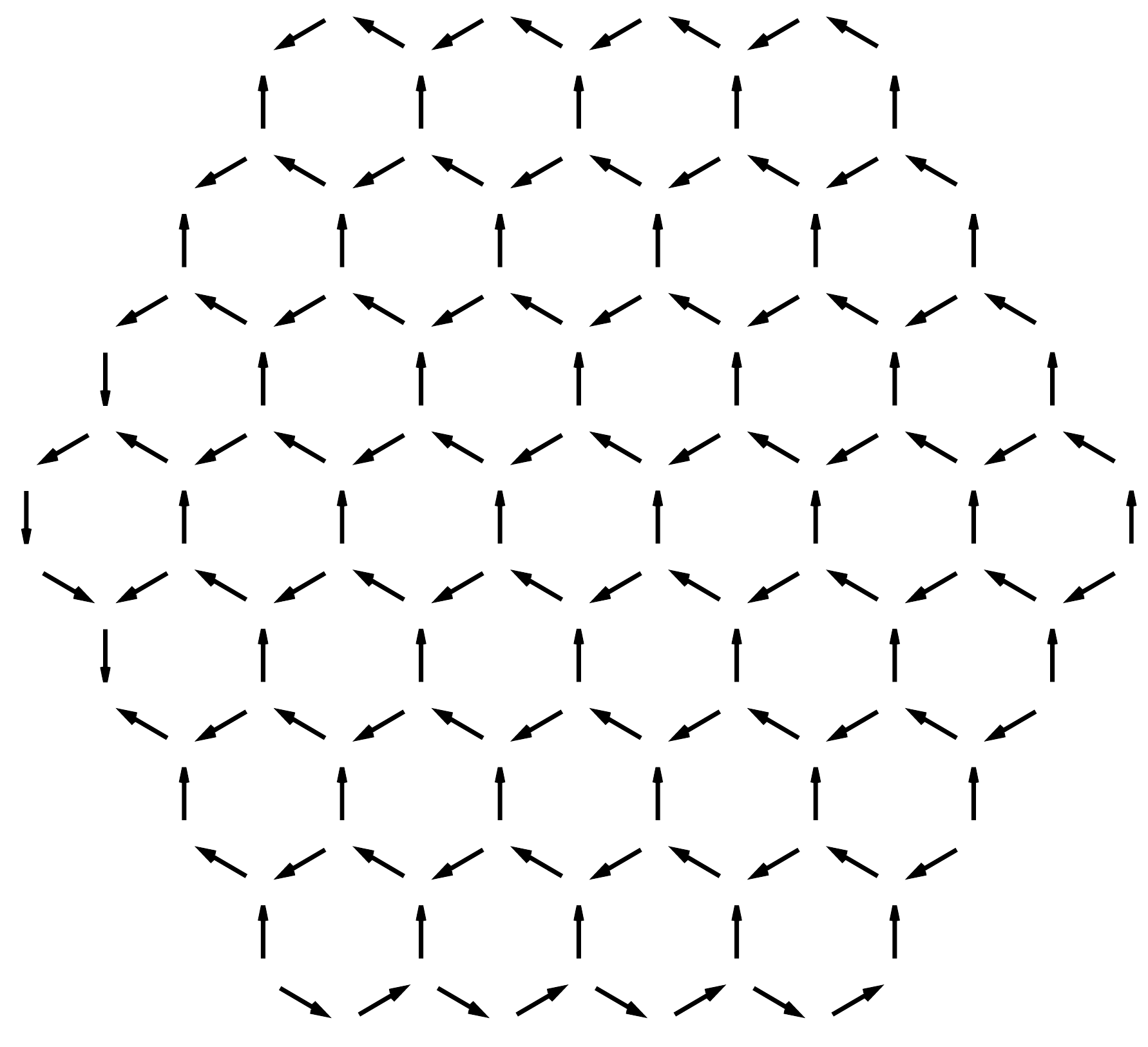}}
\caption{Schematic view of a hexagonal spin ice sample in
the form of a hexagon.} 
\label{Hex_sample}
\end{figure}

\begin{figure}
\includegraphics[width=1\linewidth]{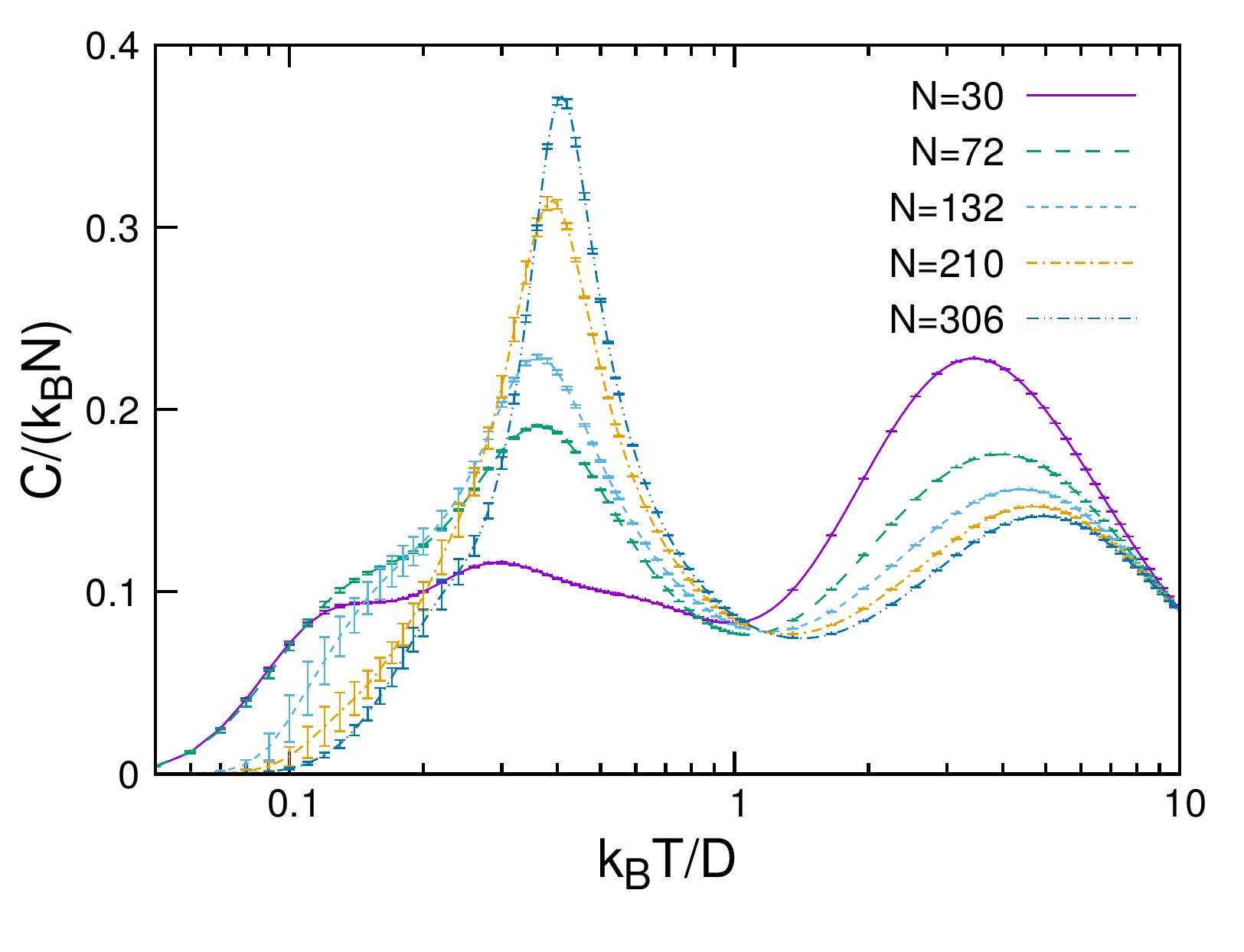}
\caption{Specific heat as a function of temperature for hexagonal spin ice (hexagonal sample) for LR
interaction.}
\label{Hex_LR}
\end{figure}

\begin{figure}
\includegraphics[width=1\linewidth]{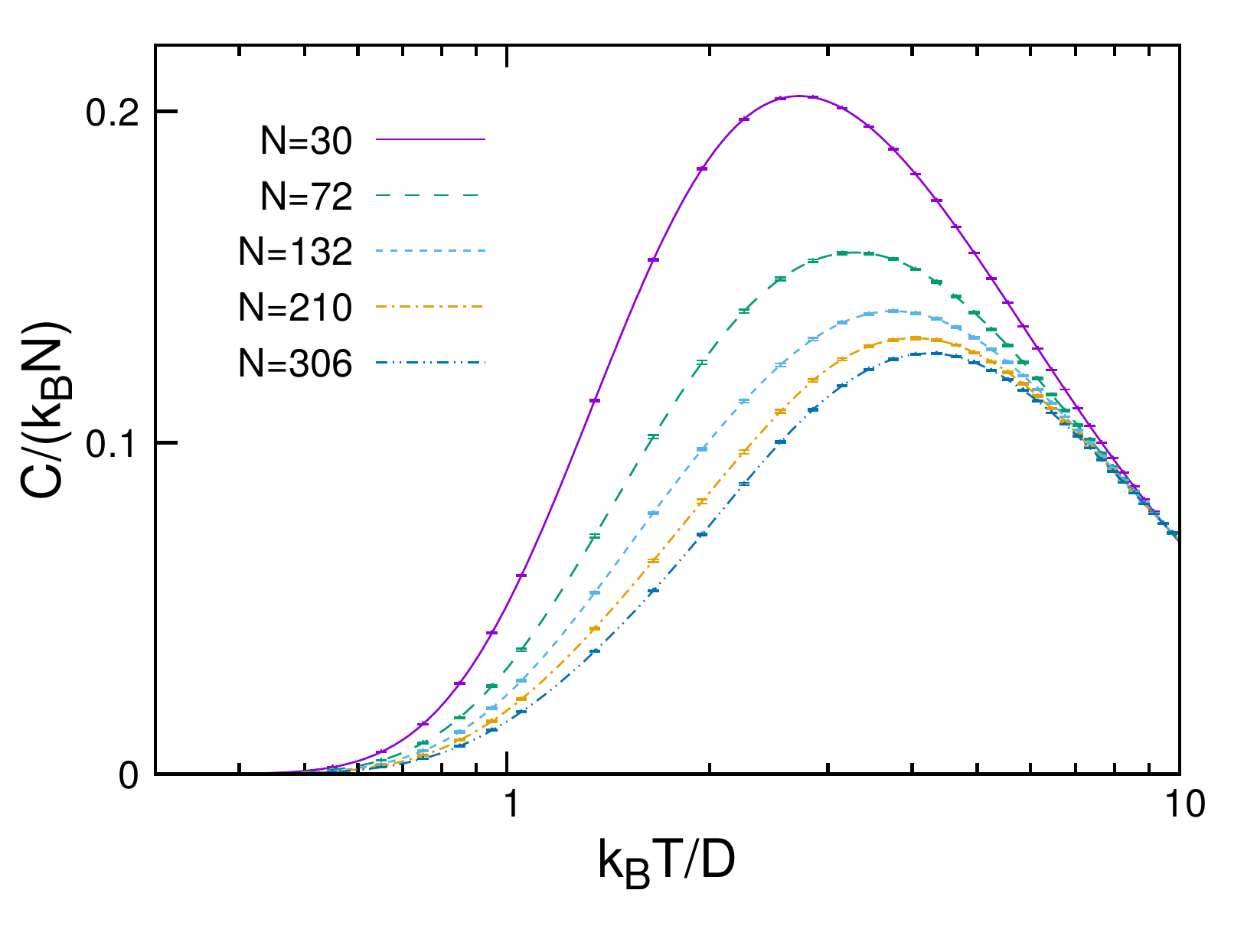}
\caption{Specific heat as a function of temperature for hexagonal spin ice (hexagonal sample) for SR
interaction.}
\label{Hex_SR}
\end{figure}

The model of spin ice for a hexagonal sample of
hexagonal lattice with free boundary conditions and
dipole interaction (\ref{dipint1}) is schematically demonstrated in Fig. \ref{Hex_sample}. The temperature behavior of the specific
heat of samples of this shape is shown in Figs. \ref{Hex_LR} and \ref{Hex_SR}.

\begin{figure}
\includegraphics[width=1\linewidth]{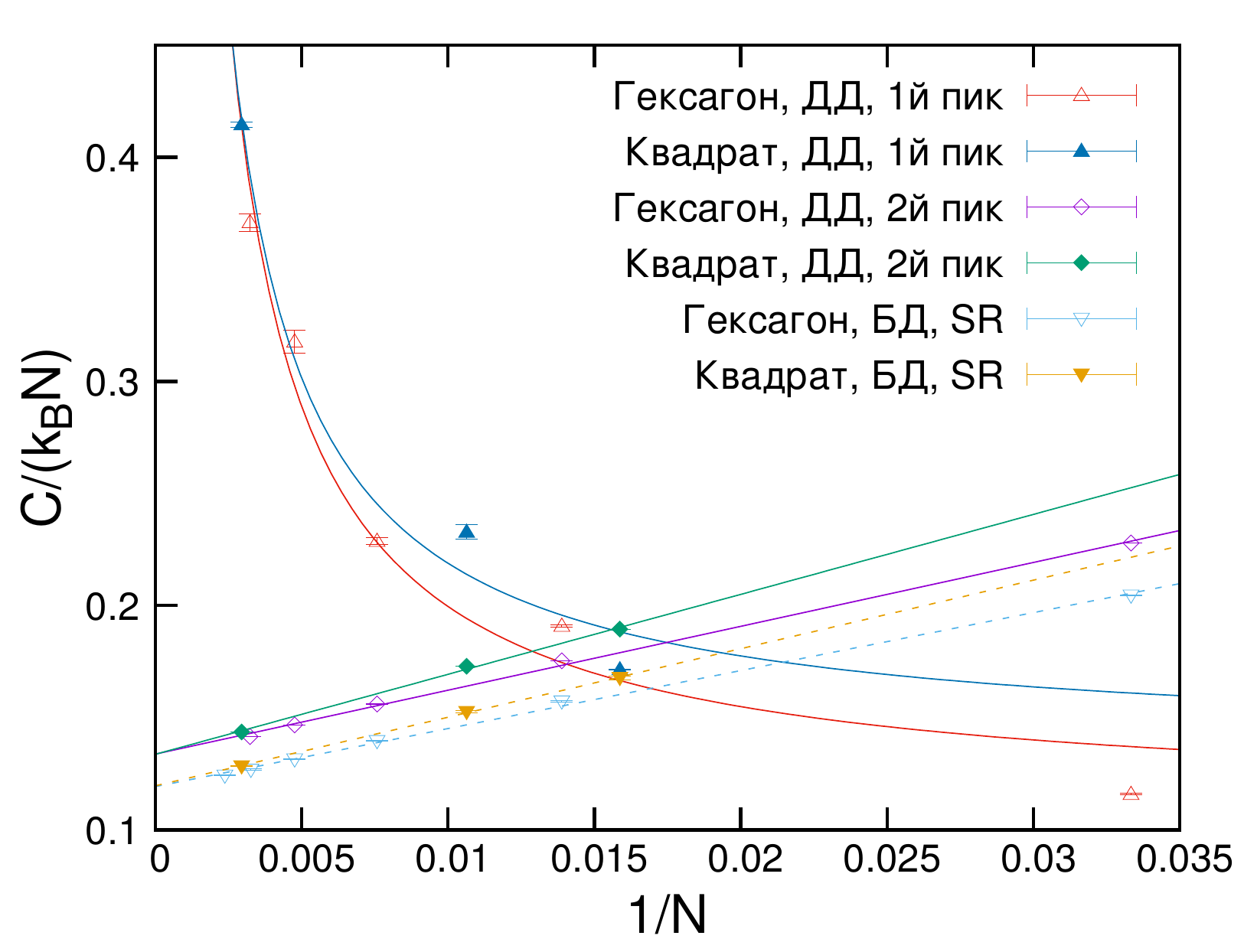}
\caption{
Comparison of the growth of specific heat peaks depending on the number of particles in
the system for hexagonal lattice models of square and hexagonal samples. As $N\rightarrow\infty$, for LR interaction, the temperature behavior of specific heat has a singularity, and the
second peak tends to$0.133$ $D/k_B$, while, for SR interaction,
the specific heat peak tends to $0.119$ $D/k_B$.
}
\label{Compare_hexagon_lattice}
\end{figure}

The values of specific heat peaks for square and
hexagonal samples of hexagonal lattice are shown in
Fig. \ref{Compare_hexagon_lattice} as a function of the size of the system. We constructed approximations by the following equations:

1) SR (square) $3.05759/N + 0.11964$, 

2) SR (hexagonal) $2.58662/N + 0.11929$,

3) LR (square) second peak $3.56674/N + 0.13368$,

4) LR (hexagonal) second peak $2.84817/N + 0.13379$,

5) LR (square) first peak $0.00083 N + 0.13619$,

6) LR (hexagonal) first peak $0.00089 N + 0.11031$.

Figure \ref{Compare_hexagon_lattice} implies that, as $N\rightarrow\infty$, one can neglect
the effect of boundaries on the critical phenomena
near $T_c$. The specific heat peaks converge in square
and hexagonal samples. In the SR case, the height of
the specific heat peak tends to a value of $0.119$ $D/k_B$; in
the LR case, the specific heat has a singularity in a
low-temperature region, and the height of the second
peak tends to a value of$0.133$ $ D/k_B$ as the size of the
system increases. This implies that there is no phase
transition in the model with SR, whereas, in the model
with LR, a phase transition occurs. The question of
how many neighbors should one take into account in order not to lose the main thermodynamic properties
of the system requires additional analysis.

\subsubsection{Spin snow}
Magnetic nanoparticles provide a
truly inexhaustible variety of many-body systems for
both experimenters and theoreticians, because the
possibilities of designing nanoparticle arrays of arbitrary architecture are actually infinite. A free choice of
the geometry of an array, as well as the geometry of its
nanocomponents, allows the design of any lattices that
have no analogs in nature. We found an analogy with
an infinite variety of snowflakes of ordinary water ice
that form ordinary snow. The present section is titled
“spin snow,” because by “snow” one usually means a
set of ice crystals. In this section, as an example, we
demonstrate the possibilities of simulation of arbitrary
samples of spin snow with a given number of neighbors, i.e., with arbitrary interaction radius.

\begin{figure}
\begin{minipage}[h]{0.34\linewidth}a)
    \includegraphics[width=1\linewidth]{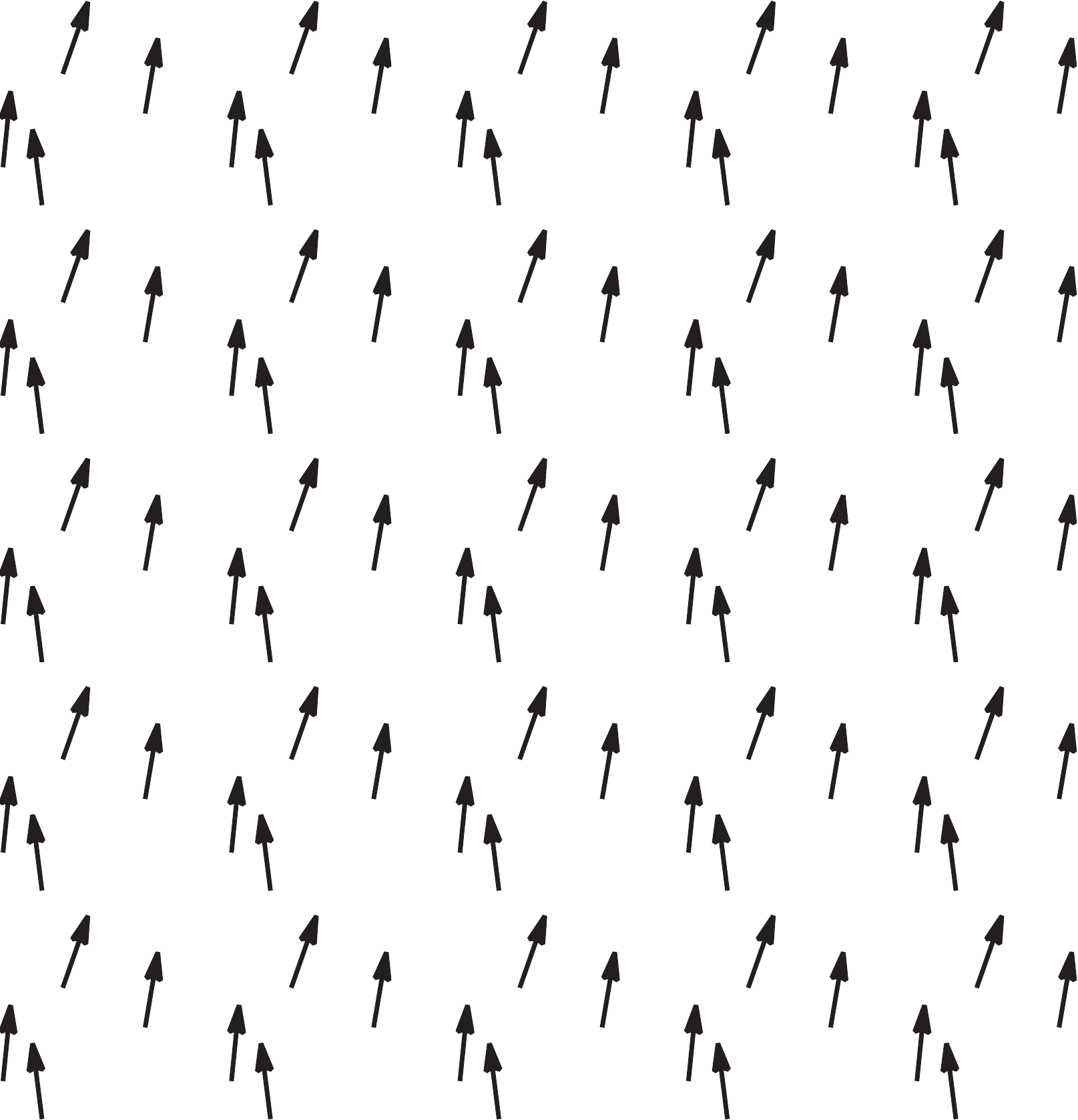}
\end{minipage}
\hfill
\begin{minipage}[h]{0.63\linewidth}
    \includegraphics[width=1\linewidth]{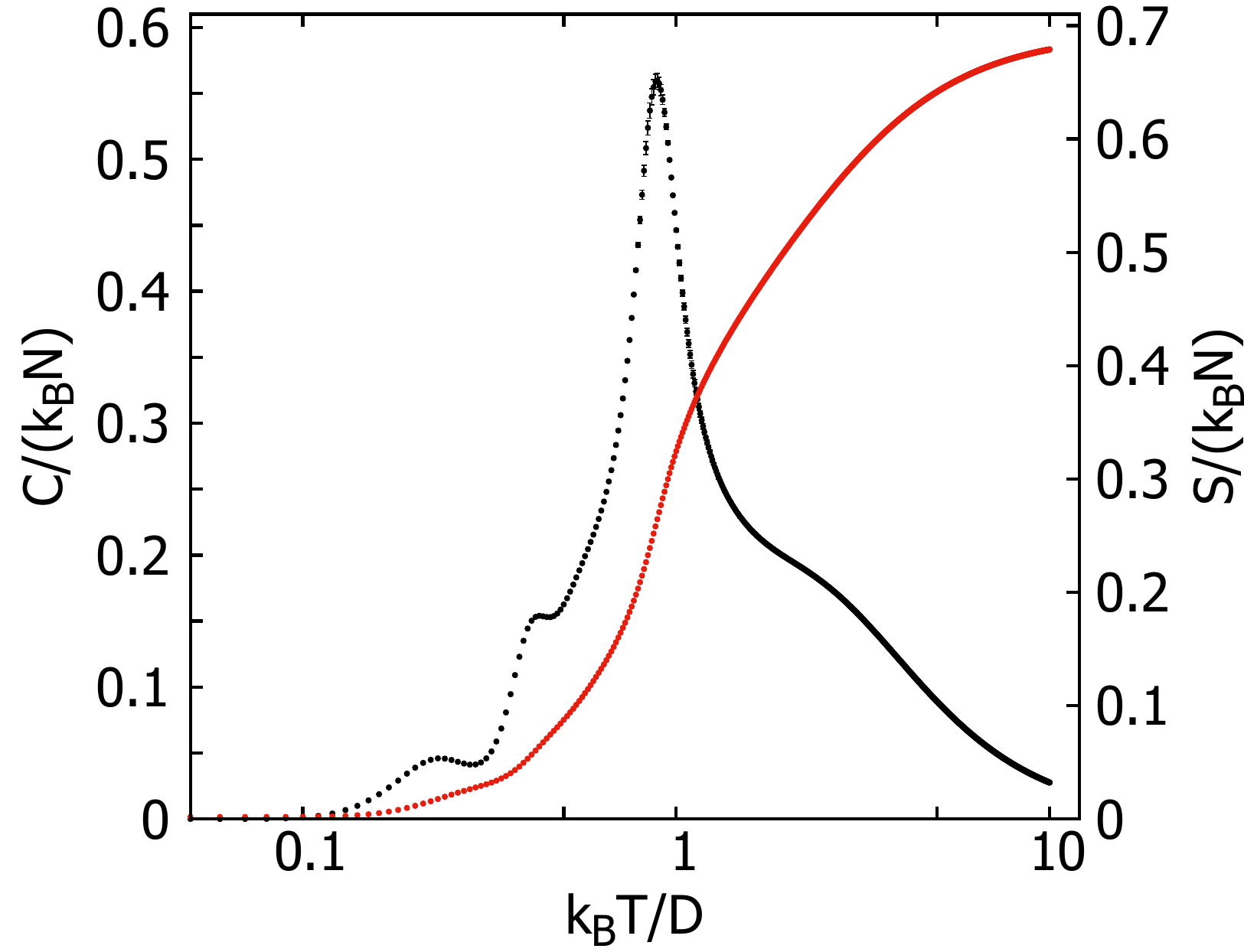}
\end{minipage} \\
\vfill
\begin{minipage}[h]{0.34\linewidth}b)
    \includegraphics[width=1\linewidth]{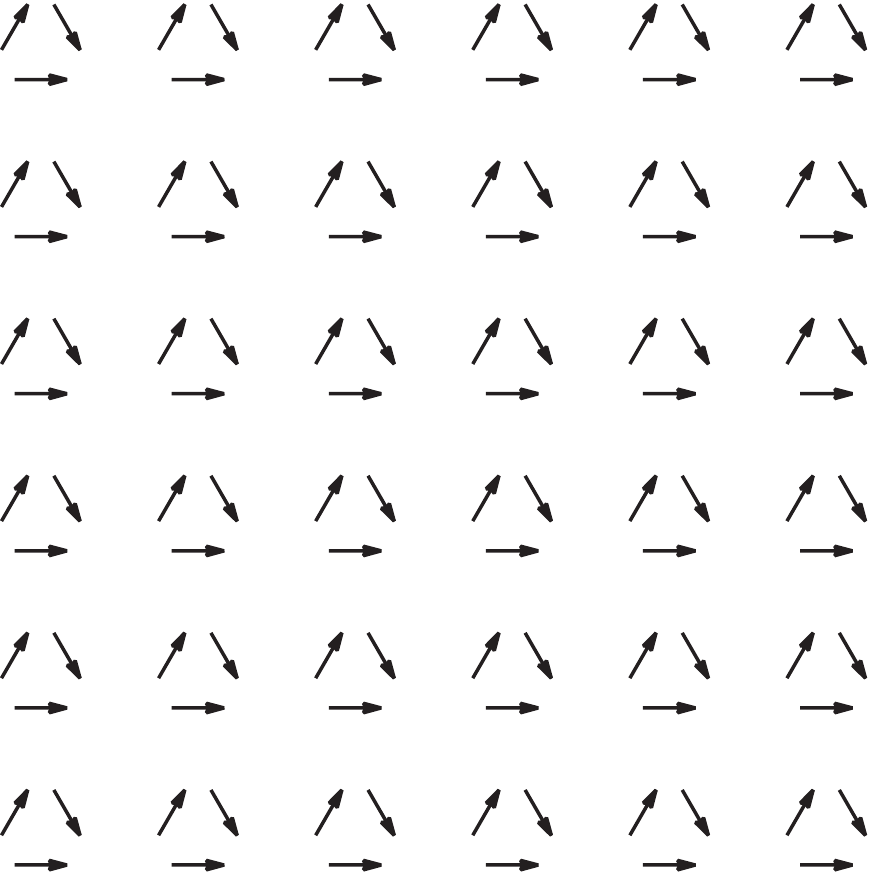}
\end{minipage}
\hfill
\begin{minipage}[h]{0.63\linewidth}
    \includegraphics[width=1\linewidth]{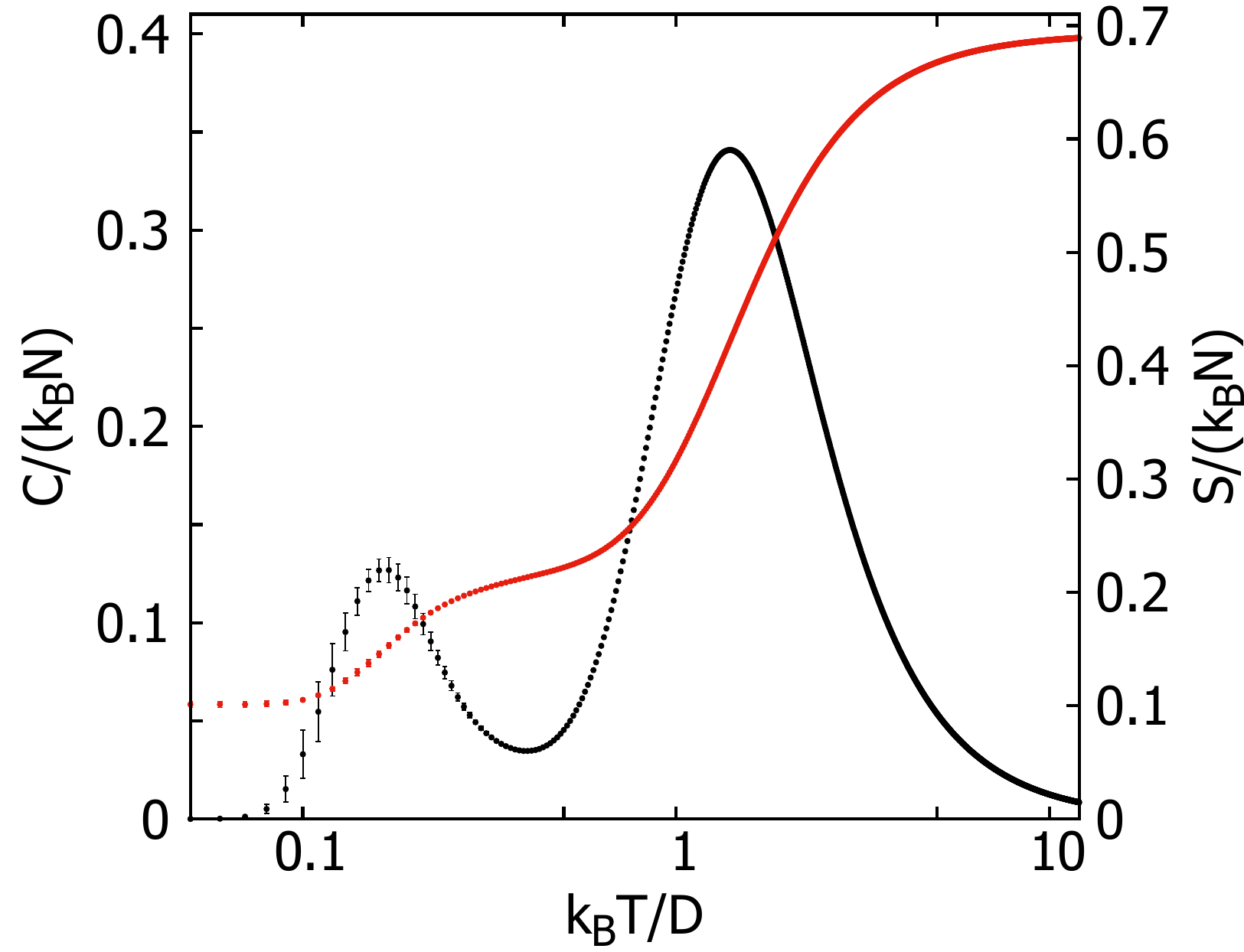}
\end{minipage} \\
\vfill
\begin{minipage}[h]{0.34\linewidth}c)
    \includegraphics[width=1\linewidth]{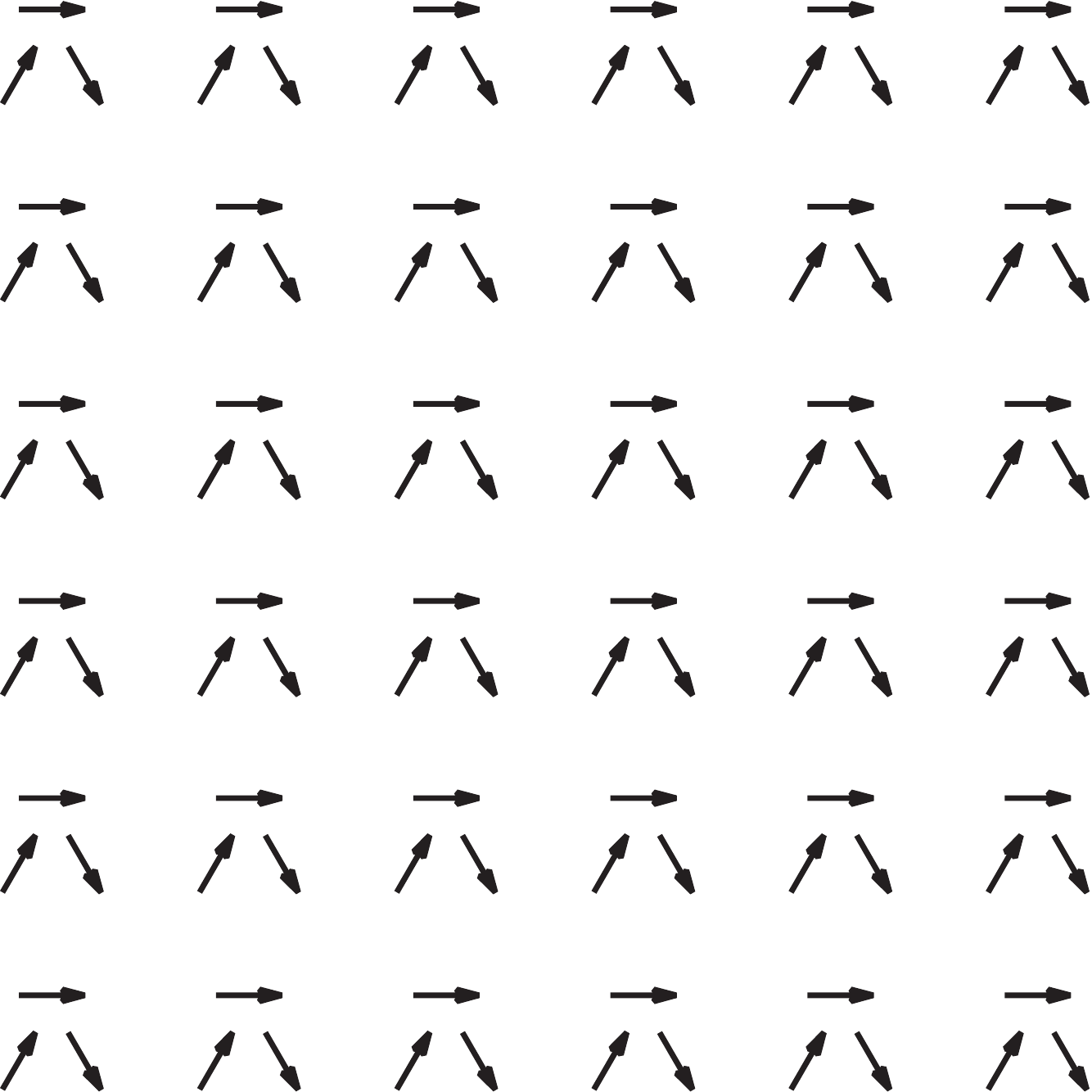}
\end{minipage}
\hfill
\begin{minipage}[h]{0.63\linewidth}
    \includegraphics[width=1\linewidth]{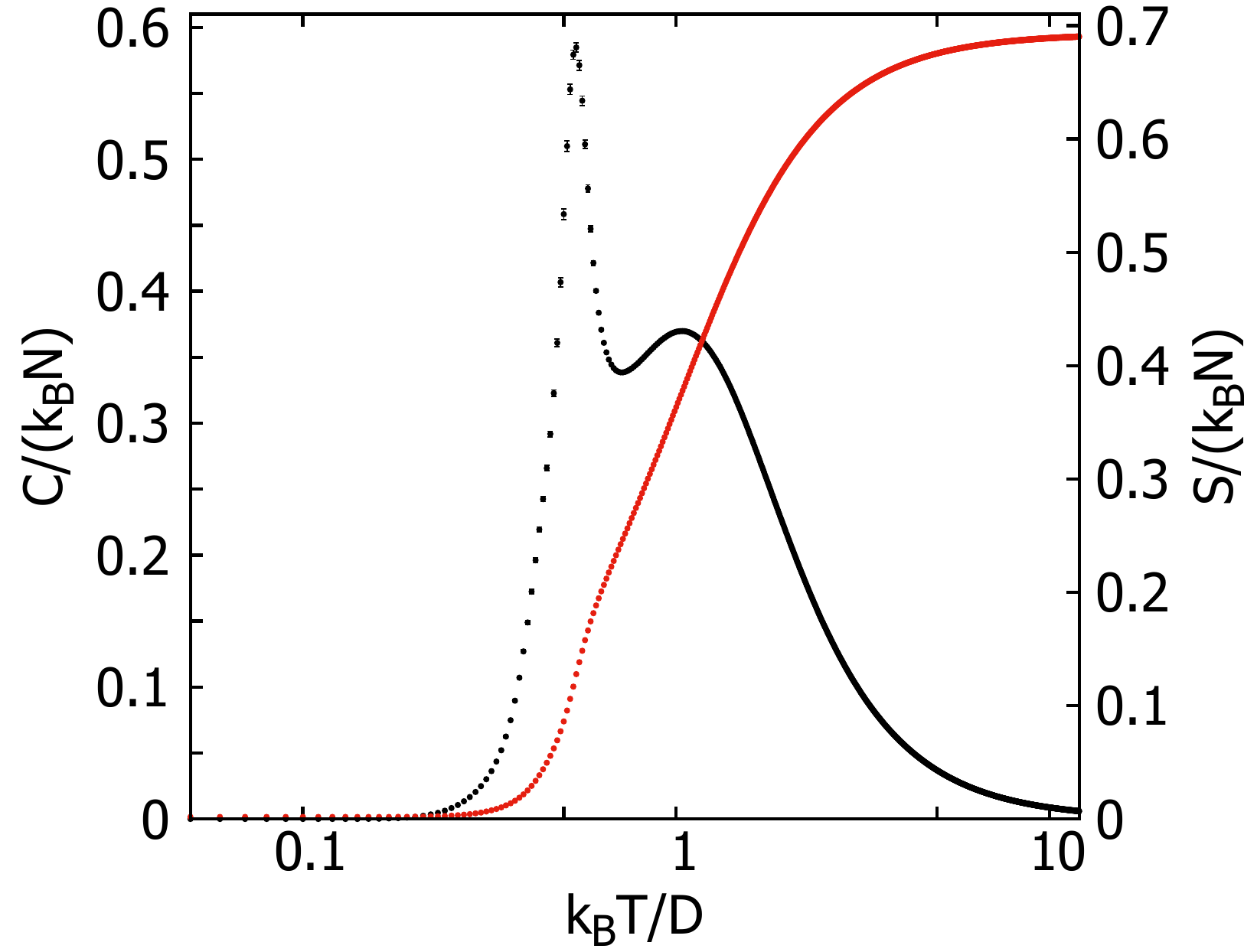}
\end{minipage}
  \caption{Thermodynamics of spin snow. (Left) schematic view of the quasilattice and (right) specific heat and
entropy as a function of temperature.}
  \label{snow}
\end{figure}

In particular, we analyzed the quasilattices shown
in Fig. \ref{snow} by a serial WL method. The calculations
were performed with regard to different coordination
spheres, depending on which the number of neighbors
in various samples ranged from 14 to 57. It turns out
that such lattices exhibit very interesting thermodynamic properties.

For example, in a sample consisting of 400 spins
(Fig. \ref{snow}a), the radius of the coordination sphere was
taken equal to four lattice constants. Therefore, the
number of neighbors ranged from 18 to 20 depending
on the location of the spins. It is noteworthy that the
temperature behavior of specific heat exhibits three
finite peaks.
The graph of specific heat of a sample consisting of
108 spins (Fig. \ref{snow}b) has two characteristic global
peaks. The radius of the coordination sphere was taken
equal to six lattice constants, and the number of neighbors ranged from 37 to 57. We observed a nonzero
value of residual entropy $S(T\rightarrow 0)$.

The temperature behavior of the specific heat of a
spin snow lattice consisting of 432 spins (Fig. \ref{snow}c) has
two peaks. The interaction between dipoles was
restricted to four coordination spheres, which
amounted to from 14 to 16 neighbors per particle.

\section{CONCLUSIONS}

In this study, we have considered serial and parallel
implementations of the WL algorithm. The scheme of
numerical calculation of the DOS proposed here involves random walks over states, and the transition
probability between states depends on the ratio of the
occurrence frequency of a previous state to the occurrence frequency of a new state. Thus, the WL method
suggests more frequent visit of energy levels with
smaller value of $g(E)$. The method shows good convergence of results to the exact solution of the twodimensional Ising model, as well as convergence to the
results of independent simulation of the SSI model by
the Metropolis method.

The divergence of specific heat for $T \gtrsim T_{c}$ in Fig. \ref{ssi_wl}
is due to an insufficient number of Metropolis steps.

A parallel WL algorithm has many variable parameters, such as $f_{min}$, $d_{min}$, $\Delta E$, $h$, $m$, and the size and
overlapping of energy windows, which affect the speed
and accuracy of calculations. However, currently there
does not exist a universal algorithm for choosing these
parameters. Fine tuning of the algorithm is performed
depending on the complexity of the energy landscape
and the type of the system considered.

The anomalous behavior of specific heat observed
in the arrays should be investigated in detail to find out
the reasons behind the variation of critical behavior
near the critical temperature of phase transition. Here
two directions of research are possible that are aimed
at the theoretical description of the second-order
phase transition.

The first direction can obviously be related to the
fact that the classical idea of the second-order phase
transition suggests mutual transformation between
two phases, which is usually explained in terms of percolation theory, i.e., in terms of the emergence of a
“percolation cluster.” The presence of one or several
peaks near $T_c$ can be attributed to the existence of several transitions between possibly coexisting phases and
the creation and subsequent transformation of a percolation cluster under LR dipole interaction. The
question of the concept of percolation cluster in systems with alternating-sign LR interaction also remains
open.

The second direction of research can be associated
with the determination of the critical behavior, which
is no longer described within universal and simple
power laws; i.e., the hypothesis of the universality of
the critical behavior of second-order phase transitions
within one of $\hatH(2,n)$-vector models requires verification.

\section*{Acknowledgments}
We are grateful to Far Eastern Federal University
for providing supercomputer facilities and to Kseniya
Valer’evna Shapovalova for preparing interesting samples of spin snow. We are also grateful to Prof. Yutaka
Okabe for valuable and helpful discussions.
This work was supported by a grant from the President of the Russian Federation for young scientists
and graduate students within the program for the development of the priority direction “Strategic Information Technologies, Including the Creation of
Supercomputers and Software Development,” project
nos. SP-946.2015.5 and SP-1675.2015.5, and under
state task “Magnetic Properties Multiscale Structure
of Nanomaterials” (task no. 3.7383.2017/B4).


\begin{thebibliography}{99}

\bibitem{Wales2003}
T. F. Middleton and D. J. Wales, J. Chem. Phys. {\bf 118}, 4583 (2003).

\bibitem{PhysRevLett.86.2050}
F. Wang and D.~P. Landau, Phys. Rev. Lett. {\bf 86}, 2050 (2001)

\bibitem{Landau2004}
D. P. Landau, S. H. Tsai and M. Exler, Am. J. Phys. {\bf 72}, 1294 (2004).

\bibitem{PhysRevE.64.056101}
F. Wang and D. P. Landau, Phys. Rev. E {\bf 64}, 056101 (2001).

\bibitem{doi:10.1142/S0129183102003243}
B. J. Schulz, B. Kurt and M. M{\"u}ller, Int. J. Mod. Phys. C  {\bf 13}, 477 (2002).
  
\bibitem{ref1}
A. Proykova and D. Stauffer, Open Phys. {\bf 3}, 209 (2005).

\bibitem{PhysRevLett.90.120201}
M. Troyer, S. Wessel and  F. Alet, Phys. Rev. Lett. {\bf 90}, 120201 (2003).

\bibitem{PhysRevE.70.066128}
A. Malakis, A. Peratzakis and N. G. Fytas, Phys. Rev. E {\bf 70}, 066128 (2004).

\bibitem{Brown2005}
G. Brown and T. C. Schulthess, J. Appl. Phys. {\bf 97}, 10E303 (2005).

\bibitem{PhysRevE.71.046705}
B. J. Schulz, K. Binder and M. M{\"u}ller, Phys. Rev. E {\bf 71}, 046705 (2005).

\bibitem{PhysRevE.72.056710}
S. Reynal and H. T. Diep, Phys. Rev. E {\bf 72}, 056710 (2005).

\bibitem{PhysRevE.82.046703}
F. Calvo, Phys. Rev. E {\bf 82}, 046703 (2010).

\bibitem{PhysRevE.89.013311}
Y.~L. Xie, P.~Chu, Y.~L. Wang et al., Phys. Rev. E {\bf 89}, 013311 (2014).

\bibitem{Calvo2003}
F.~Calvo and P.~Parneix, J. Chem. Phys. {\bf 119}, 256 (2003).

\bibitem{PhysRevB.72.214203}
J. Snider and C. Y. Clare, Phys. Rev. B {\bf 72}, 214203 (2005).

\bibitem{PhysRevLett.110.210603}
T. Vogel, Y. W. Li, T. W{\"u}st et al, Phys. Rev. Lett. {\bf 110}, 210603 (2013).

\bibitem{PhysRevE.72.036702}
D. Jayasri, V. S. S. Sastry and K. P. N. Murthy, Phys. Rev. E {\bf 72}, 036702 (2005).

\bibitem{Desgranges2009}
C. Desgranges and J. Delhommelle, J. Chem. Phys. {\bf 130}, 244109 (2009).

\bibitem{Ngo2010}
V. T. Ngo, D. T. Hoang and H. T. Diep, Phys. Rev. E {\bf 82}, 041123 (2010).

\bibitem{PhysRevE.92.022134}
W. Kwak, J. Jeong, J. Lee et al., Phys. Rev. E {\bf 92}, 022134 (2015).

\bibitem{Caparica2015447}
AA Caparica, S. A. Le{\~a}o and C. J. DaSilva, Physica A: Statistical Mechanics and its Applications {\bf 438}, 447 (2015).

\bibitem{PhysRevE.90.042715}
J. Liu, B. Song, Y. Yao et al., Phys. Rev. E {\bf 90}, 042715 (2014).

\bibitem{Rathore2002}
N. Rathore and J. J. de Pablo, J. Chem. Phys. {\bf 116}, 7225 (2002).

\bibitem{TSAI2006}
S. H. Tsai, F. Wang and D. P. Landau, Brazilian journal of physics {\bf 36}, 635 (2006).

\bibitem{ArgollodeMenezes2003428}
M. A. De Menezes and A. R. Lima, Physica A: Statistical Mechanics and its Applications {\bf 323}, 428 (2003).

\bibitem{0305-4470-36-24-304}
V. Mustonen and R. Rajesh, J. Phys. A: Math. Gen. {\bf 36}, 6651 (2003).

\bibitem{PhysRevE.72.025701}
C. Zhou and R. N. Bhatt, Phys. Rev. E {\bf 72}, 025701 (2005).

\bibitem{PhysRevE.75.046701}
R. E. Belardinelli and V. D. Pereyra, Phys. Rev. E {\bf 75}, 046701 (2007).

\bibitem{Metropolis1953}
N. Metropolis, A. W. Rosenbluth, M. N. Rosenbluth et al., J. Chem. Phys. {\bf 21}, 1087 (1953).

\bibitem{PhysRevLett.58.86}
R. H. Swendsen and J. S. Wang, Phys. Rev. Lett. {\bf 58}, 86 (1987).

\bibitem{earl2005parallel}
D. J. Earl and M. W. Deem, Phys. Chem. Chem. Phys. {\bf 7}, 3910 (2005).

\bibitem{PhysRevLett.62.361}
U. Wolff, Phys. Rev. Lett. {\bf 62}, 361 (1989).

\bibitem{Bornn2013749}
L.~Bornn, P. E. Jacob and P.~Del~Moral, J. Comput. Graph. Stat. {\bf 22}, 749 (2013).

\bibitem{Bauer2010}
B.~Bauer, E.~Gull, S.~Trebst et al., Journal of Statistical Mechanics: Theory and Experiment {\bf 2010}, No. P01020 (2010).

\bibitem{Xu2015}
S.~Xu, X.~Zhou, Y.~Jiang et al., Sci. China Phys., Mech. Astron., {\bf 58}, 1 (2015).

\bibitem{Belardinelli2008}
R. E. Belardinelli, S.~Manzi and V. D. Pereyra, Phys. Rev. E {\bf 78}, 067701 (2008).

\bibitem{Maerzke2012}
K. A. Maerzke, L.~Gai, P. T. Cummings et al., J. Chem. Phys. {\bf 137}, 204105 (2012).

\bibitem{PhysRevLett.92.097201}
P.~Dayal, S.~Trebst, S.~Wessel et al., Phys. Rev. Lett. {\bf 92}, 097201 (2004).

\bibitem{Belardinelli2007}
R. E. Belardinelli and V. D. Pereyra, J. Chem. Phys., {\bf 127}, 184105 (2007).

\bibitem{wang2006artificial}
R. F.~Wang, C.~Nisoli, R. S.~Freitas et al., Nature {\bf 439}, 303 (2006).

\bibitem{PhysRevB.77.094418}
Y.~Qi, T.~Brintlinger and J. Cumings, Phys. Rev. B {\bf 77}, 094418 (2008).

\bibitem{Lederman1993}
M.~Lederman, G.~A. Gibson and S.~Schultz, J. Appl. Phys. {\bf 73}, 6961 (1993).

\bibitem{PhysRevB.80.140409}
G.~M{\"o}ller and R.~Moessner, Phys. Rev. B {\bf 80}, 140409 (2009).

\bibitem{Nefedev2010}
K.~V. Nefedev, Y.~P. Ivanov and A.~A. Peretyatko, Methods and Tools of Parallel Programming Multicomputers: Second Russia-Taiwan Symposium, MTPP 2010, Vladivostok, Russia, May 16-19, 2010, Revised Selected Papers {\bf 6083}, 260 (2010).

\bibitem{Ivanov2011}
Y.~P. Ivanov, K.~V. Nefedev, A.~I. Iljin et al., Journal of Physics: Conference Series {\bf 266}, 012117 (2011).

\bibitem{Nefedev2011}
K. V. Nefedev, Y. P. Ivanov, A. A. Peretyatko  et al., Solid State Phenomena {\bf 168}, 325 (2011).

\bibitem{RevModPhys.85.1473}
C. Nisoli, R. Moessner and P. Schiffe, Rev. Mod. Phys. {\bf 85}, 1473 (2013).

\bibitem{landau2000guide}
D. P Landau and K. Binder, Cambridge Univ. Press, 384 (2000).

\bibitem{PhysRevE.84.065702}
G.~Brown, Kh. Odbadrakh, D.~M. Nicholson  et al., Phys. Rev. E {\bf 84}, 065702 (2011).

\bibitem{kalyan2016joint}
M. S. Kalyan, R. Bharath, V. S. S. Sastry  et al., J. Stat. Phys. {\bf 163}, 197 (2016).

\bibitem{silant2011}
I. A. Silant’eva and P. N. Vorontsov-Vel’yaminov, Vychisl. Metody Programm. {\bf 12}, 397 (2011).

\bibitem{shchure2014}
L. N. Shchur, Mekh., Upravl. Inform. {\bf 6} (6), 160 (2014).

\bibitem{vogel2014scalable}
T. Vogel, Y. W. Li, T. W{\"u}st et al., Phys. Revi. E {\bf 90}, 023302 (2014).

\bibitem{silva2012thermodynamics}
R. C. Silva, F. S. Nascimento, L. A. S. M{\'o}l et al., New Journal of Physics {\bf 14}, 015008 (2012).

\bibitem{Vaz2008}
C.A.F. Vaz, J.A.C. Bland, G. Lauhoff, Reports on Progress in Physics, {\bf 71},  056501
(2008).

\bibitem{Fisher1974} M.E. Fisher, Rev. Mod. Phys. {\bf 46} 597 (1974).

\bibitem{Wilson1979} K.G. Wilson, Sci. Am. {\bf 241} 140 (1979).

\bibitem{Fisher1998} M.E. Fisher, Rev. Mod. Phys. {\bf 70} 653 (1998).

\bibitem{Stanley1999} E.E. Stanley, Rev. Mod. Phys. {\bf 71} S358 (1999)

\bibitem{Ma1976} S.K. Ma, Modern Theory of Critical Phenomena (Reading, MA: Benjamin), (1976).

\bibitem{Jose1977}  J.V. Jose, L.P. Kadanoff, S. Kirkpatrick and D.R. Nelson, Phys. Rev. B {\bf 16} 1217 (1977).

\bibitem{Kosterlitz1978}  J.M. Kosterlitz and D.J. Thouless, Progress in Low Temperature Physics vol. VII B ed. D.F. Brewer
(Amsterdam: North-Holland) p. 371  (1978).

\bibitem{Shevchenko2017} Y. Shevchenko, A. Makarov, K. Nefedev, Phys. Let. A, {\bf 381} 428 (2017).

\bibitem{Ferdinand1969} A.E. Ferdinand and M.E. Fisher, Phys. Rev. B {\bf 185}, (832) 1969.


\end{thebibliography}
\end{document}